\documentclass{aa}
\usepackage{amssymb,amsmath}
\usepackage{txfonts}
\usepackage[colorlinks=true]{hyperref}
\hypersetup{
    colorlinks=true,
    linkcolor=blue,
    citecolor=blue,
    urlcolor=blue
}

\newcommand{\omegam}{\Omega^{-1}}

\newcommand{\source}{{\cal S^+} }
\newcommand{\sink}{{\cal S^-} }
\newcommand{\peff}{p_{\rm eff}}
\newcommand{\st}{{\rm St}}
\newcommand{\ey}{{\bf \hat y}}
\newcommand{\ex}{{\bf \hat x}}
\newcommand{\ugas}{{\bf  u}}
\newcommand{\vdust}{{\bf  v}}
\newcommand{\dvpp}{\Delta v_{\rm pp}}
\newcommand{\alphass}{\alpha_{\rm SS}}

\begin{document}
\title{Coexistence of  coagulation and streaming instabilities in protoplanetary discs}
\titlerunning{Coagulation and Streaming instabilities} 
\author{Arnaud Pierens 
 \inst{1}
 \and
  Thomas Collin-Dufresne \inst{1}
  \and
  Min-Kai Lin  \inst{2,3} 
  \and 
  Emmanuel DiFolco \inst{1}}
\institute{ Laboratoire d'astrophysique de Bordeaux, Univ. Bordeaux, CNRS, B18N, all\'ee Geoffroy Saint-Hilaire, 33615 Pessac, France\\
\email{arnaud.pierens@u-bordeaux.fr}
\and
  Institute of Astronomy and Astrophysics, Academia Sinica, Taipei 10617, Taiwan
  \and 
  Physics Division, National Center for Theoretical Sciences, Taipei City, 10617, Taiwan
  }

\abstract{
The streaming instability is considered one of the leading candidates for the formation of planetesimals, due to its ability to overcome the bouncing and fragmentation barriers. The formation of dense dust clumps through this process, however, is possible provided it involves solids with dimensionless stopping times $\sim 0.1$ in standard discs, which typically corresponds to 1-10 cm-sized particles. This implies that dust coagulation is required for the SI to be an efficient process. Here, we employ unstratified, shearing-box simulations combined with a moment equation for solving the coagulation equation to examine the effect of dust growth on the SI. In dust-rich discs with a dust-to-gas ratio $\epsilon\gtrsim 1$, coagulation is found to have little impact on the SI;  while in dust-poor discs with $\epsilon\sim 0.01$,  we observe the formation of vertically extended filaments through the action of the coagulation instability (CI), which is triggered due to the  dependence of coagulation efficiency on dust density. For moderate dust-to-gas ratios $\epsilon\sim 0.1$ and Stokes numbers $\st \lesssim 0.1$,  we find  onset of the SI within these filaments, with a linear growth rate significantly higher compared to standard SI. We refer to this regime as 
coagulation-assisted SI. The synergy between both instabilities in that case leads to isotropic turbulence and dust concentrations that are increased by a factor of $30-40$.  As dust continues to grow, SI tends to overcome the effect of the CI such that the nonlinear saturation phase is similar to pure SI. Our results  suggest that coagulation, by simply increasing dust size,  may facilitate the formation of dense clumps through the SI; even though it has only little effect on its nonlinear evolution.  
}
\keywords{
accretion, accretion discs --
                planet-disc interactions--
                planets and satellites: formation --
                hydrodynamics --
                methods: numerical
}

\maketitle

\section{Introduction}
In the standard scenario for planet formation, coagulation of micron-sized grains leads to the formation of millimeter-sized dust  \citep{2005A&A...434..971D}. Further growth of particles above the millimeter-size range is however rendered difficult because of the bouncing \citep{2010A&A...513A..57Z} and fragmentation \citep{2010A&A...513A..57Z} barriers. An emerging picture to overcome these  barriers is that 100-km sized planetesimals form directly through the streaming instability (hereafter SI; \citet{2005ApJ...620..459Y,2007ApJ...662..613Y}) that occurs as a result of  angular momentum exchange between the gas and dust components, and rotation.  It has been shown that during the nonlinear evolution of the SI, dust particles can  eventually directly concentrate into clumps or filaments, which can  subsequently become gravitationally unstable to form 100-1000 km-sized bodies \citep{2009ApJ...704L..75J,2016ApJ...822...55S,2019ApJ...883..192A}. The ability for solids to experience a regime of strong clumping depends mainly on the dimensionless   stopping time $\st$ and local dust-to-gas ratio $\epsilon$, which can also be quantified through the dust-to-gas surface density ratio $Z=\Sigma_d/\Sigma_g$ where $\Sigma_d$ (resp. $\Sigma_g$) is the dust (resp. gas) surface density.  For $Z\sim 5\times 10^{-3}$,  strong clumping requires $\st \gtrsim 0.1$  in a disc where no external turbulence operates, whereas for $\st\lesssim 0.01$ strong clumping is triggered for a critical solid abundance of a few percent \citep{2015A&A...579A..43C,2017A&A...606A..80Y,2021ApJ...919..107L}.  In any case, this suggests that a level of dust growth is needed for the SI to enter the strong clumping regime \citep{2014A&A...572A..78D}.  

Taking into account the effect of dust growth on the SI in a self-consistent fashion is  however a difficult task. The  process of dust growth is indeed expected to be a sensitive function of  the relative collision velocity between grains,  which depends on many processes operating in the protoplanetary disc. As a consequence, the resulting dust size distribution is not known a priori and remains uncertain.  The generalization of the classical, monodisperse SI to multiple  dust sizes have nevertheless been examined by  a number of recent studies, but adopting prescribed dust size distributions. These have shown that the growth rate of the polydisperse, multispecies SI (PSI) tends to decrease as the number of species is increased \citep{2019ApJ...878L..30K}, and becomes very small in the limit of a continuous size distribution \citep{2020MNRAS.499.4223P,2021MNRAS.501..467Z}.  However, it was also found that a top-heavy size distribution, resulting for instance from a pure coagulation process, leads to PSI linear growth rates equivalent to those of the classical SI \citep{2021MNRAS.502.1469M}. 

The influence of the process of coagulation itself on the nonlinear development of the monodisperse SI has been  investigated by \citet{2024ApJ...975L..34H} using stratified simulations.  These authors found that coagulation tends to broaden the range of Stokes numbers and dust-to-gas ratios over which the SI can be triggered, due to ability for coagulation to promote dust growth. Conversely, it has even been proposed that dust growth can be significantly boosted in the strong clumping regime of the SI \citep{2023ApJ...958..168T,2025ApJ...983...15T}. This arises  because dust-loading increases the inertia of the dusty fluid, resulting in  a  reduction of the effective sound speed  and lower collision velocities between dust grains.  These results suggest that a feedback loop between coagulation and the development of the SI may exist, and this has been recently studied  by \citet{2025A&A...696L..23C}. 

In this paper, we examine the effect of the coagulation instability (hereafter CI) on the linear growth and nonlinear saturation phase of the SI \citep{2021ApJ...923...34T}.  The coagulation instability is a consequence of the interplay between the dust  coagulation process and radial drift of dust particles, and is primarily related to the dependence of  mass growth rate on  dust density. As described by \citet{2021ApJ...923...34T}, if one assumes a positive dust density perturbation, the resulting increased coagulation efficiency leads to a variation of the dust size. This subsequently causes an enhanced radial drift speed which tends to amplify  the initial density perturbation, resulting in a feedback loop.  It has been shown that the nonlinear evolution of the CI can enable the Stokes number of solids to rapidly reach unity, even for relatively modest dust-to-gas ratios $\epsilon \sim 10^{-3}$ \citep{2022ApJ...937...21T,2022ApJ...940..152T}. Therefore, the CI may represent a promising mechanism to extend the parameter space (in terms of dust-to-gas ratios and Stokes numbers) over which the  SI can operate efficiently. 

To investigate the possible effect of the CI on the SI in more detail, we perform 2D axisymmetric, unstratified, shearing-box simulations in which coagulation is modelled using the single-size approximation.  Our main aims are  to i) assess whether the  CI can help the SI  enter the regime of strong clumping and ii) evaluate to what extent the turbulence driven by the CI subsequently impacts the nonlinear development of the SI. 

The paper is organized as follows.  In Sect.2, we present the governing evolution equations of the gas and dust components. We then describe in Sect. 3 the numerical setup and the initial conditions that are used in the simulations, whose results are presented in Sect. 4.  We discuss our results  in Sect.5 and present a summary of our findings in Sect. 6. 

\section{Physical model}
\subsection{Governing equations}
We model a local patch in a protoplanetary disk within the shearing box framework. A cartesian coordinate system with origin located at an arbitrary distance $R_0$ from the central star corotates with angular velocity $\Omega_0=\Omega_k(R_0)$, with $\Omega_k$ the Keplerian angular velocity. The x- and y- axes are oriented radially outward and along the orbital direction respectively, while the z-axis is directed in the vertical direction. We assume that the domain is small compared to the orbital distance so that Keplerian rotation appears as a linear shear  flow with velocity ${\bf U_k}=-\frac{3}{2}x \Omega_0 \ey$. We assume the system is axisymmetric and neglect the vertical component of stellar gravity. In these limits, the continuity and momentum equations for the gas and dust components are respectively given by:

\begin{equation}
\frac{\partial \rho_g}{\partial t}+\nabla \cdot (\rho_g \ugas)=0
\label{eq:rhog}
\end{equation}

\begin{multline}
\frac{\partial \ugas}{\partial t}+\ugas \cdot \nabla  \ugas
= 2 u_y\Omega \ex -u_x\frac{\Omega}{2} \ey
-\frac{1}{\rho_g}\nabla P + 2\eta R \Omega^2 \ex 
+ \frac{\epsilon}{\tau_s}(\vdust-\ugas)\\
+ \frac{1}{\rho_g}\nabla\cdot \mathbf{T}
\label{eq:gasmom}
\end{multline}

\begin{equation}
\frac{\partial \rho_d}{\partial t}+\nabla \cdot (\rho_d \vdust)=\nabla\cdot(D\rho_d\nabla \epsilon)
\label{eq:rhod}
\end{equation}

\begin{equation}
\frac{\partial \vdust}{\partial t}+\vdust \cdot \nabla  \vdust=2 v_y\Omega \ex -v_x\frac{\Omega}{2} \ey-\frac{1}{\tau_s}(\vdust-\ugas)
\label{eq:vdust}
\end{equation}
 where $\rho_g$ and $\rho_d$ are the gas and dust densities respectively,  and $\ugas$, $\vdust$ their velocities measured relatively to the background Keplerian shear. We adopt an isothermal equation of state $P=\rho_g c_s^2$ where $c_s=H_g \Omega_0$ is the (constant) sound speed with $H_g$  the gas pressure scale height.  In Eq. \ref{eq:gasmom}, 
 $\mathbf{T}$ is the viscous stress tensor which is given by:
  \begin{equation}
    \mathbf{T} = \rho_g \nu \left( \nabla \ugas + (\nabla \ugas)^{\dagger}
 - \frac{2}{3} I \, \nabla\!\cdot \ugas \right)  
 \end{equation}
where $\nu$ is the kinematic viscosity which is employed to model the effects of possible underlying turbulence.  Regarding the resulting turbulence-induced particle stirring, it is modeled by the source term appearing in the dust continuity equation, where $D$   is the dust diffusion coefficient which is set here to $D=\nu$. It is important to note that viscosity and diffusion are generally not included in our simulations, except in those presented in \ref{sec:appB}.

 In Eq. \ref{eq:gasmom}, the term $\propto \eta$  corresponds to an outward force acting on the gas due to a background radial pressure gradient, and determined by the dimensionless parameter: 
 \begin{equation}
 \eta=-\frac{1}{2\rho_g \Omega_0^2 R_0}\frac{\partial P}{\partial r}
 \end{equation}
 Finally, $\tau_s$ is the stopping time that we will characterize in the following in terms of the dimensionless Stokes number $\st=\tau_s \Omega$. Dust grains are assumed to be in the Epstein regime such that $\st $ is the related to the grain size $a$ through:
 
 \begin{equation}
 \st=\sqrt{\frac{\pi}{8}}\frac{\rho_i a}{\rho_g c_s}\Omega
 \label{eq:stokes}
 \end{equation}
  where $\rho_i$ is the internal density of dust grains, and $m=4\pi \rho_i a^3/3$ their mass. In this work,  we allow $m$  (and consequently $a$) to increase as a result of  dust coagulation, which we model here using a single-size approximation \citep{2016A&A...589A..15S,2021ApJ...923...34T}. This consists in following the size evolution of grains whose mass $m_p$ dominate the dust density distribution at each location. The evolution equation for the so-called peak mass $m_p$ can be obtained by considering the first-moment of the Smoluchowski equation (e.g. \citet{2021ApJ...923...34T}, Appendix A), and it reads:
 \begin{equation}
\frac{\partial m_p}{\partial t}+\vdust \cdot \nabla  (m_p)= \source(\st_p)-\sink
\label{eq:mp}
\end{equation}
 where $\source=4 \pi a_p^2 \dvpp \rho_d$ with $a_p$ the grain size corresponding to $m_p$, and $\dvpp$  the typical relative velocity between particles, is the mass growth rate resulting from coagulation, and which can generally be expressed as a function of the Stokes number $\st_p$ corresponding to $m_p$. Although dissipation effects are not included in our dynamical equations, we follow \citet{2021ApJ...923...34T} and assume that relative velocities are dominated by  the turbulence experienced by the disc.  This allows us to i) make the problem analytically tractable and ii)  identify clearly the role played by coagulation. In that case, and in the limit $\st << 1$,  we can further make use of  a simplified expression for $\dvpp$ given by  \citep{2007A&A...466..413O}:
\begin{equation}
\dvpp=\sqrt{C\st_p \alpha_{\rm coag}}c_s;
\label{eq:dvpp}
\end{equation} 
 
  where $C$ is a numerical factor set to $C=2.3$, and  $\alpha$ the dimensionless turbulent stress parameter \citep{1973A&A....24..337S}  which is here fixed to $\alpha_{\rm coag}=10^{-4}$.  In the following and  for convenience,  we will simply drop  the subscript p in the definition of $\st_p$ and  $a_p$ such that, unless otherwise stated, $ \st=\sqrt{\frac{\pi}{8}}\frac{\rho_i a}{\rho_g c_s}\Omega$ with   $ \rho_i=3m_p/4\pi a^3$. 

With respect to  the original work of \citet{2021ApJ...923...34T}, we include in the mass evolution equation Eq. \ref{eq:mp} an additional, {\it constant} sink term $\sink$  whose value is set to the initial value of $\source$. The primary aim of adding this term is to ensure that the initial chosen particle mass corresponds to an equilibrium solution to the set of Eqs.  \ref{eq:rhog}-\ref{eq:vdust}, combined with  Eq. \ref{eq:mp}. Given that the radial and vertical dependences of equilibrium quantities are ignored in the box, the  left-hand side of Eq.  \ref{eq:mp} becomes zero at steady-state, leading therefore to  $\sink=\source$. \\

 We notice that  in principle,  dust loading leads to a reduction of the sound speed by increasing the inertia of the dusty fluid. This can be expressed by defining an effective sound speed given by \citep{2017ApJ...849..129L}:
  
  \begin{equation}
  \tilde c_s=\frac{c_s}{\sqrt{1+\epsilon}}
  \end{equation}

From Eq. \ref{eq:dvpp}, a reduction in the sound speed due to dust-loading results in a decrease  of the collision velocity.  Recent work has shown that this effect may cause  rapid dust growth in the strong clumping regime of the SI,  due to the significant increase in the local dust-to-gas ratio there \citep{2023ApJ...958..168T,2025ApJ...983...15T,2025A&A...696L..23C}.  Here, we do not account for this potential important effect and simply employ the nominal sound speed when computing the collision velocity. 

\subsection{Equilibrium state}
For constant values of $\rho_g$, $\rho_d$ and $\st$, equilibrium solutions to Eqs. \ref{eq:rhog}-\ref{eq:vdust} can be obtained,  leading to velocity deviations from Keplerian rotation given by: 
\begin{equation}
u_x=\frac{2\epsilon \st}{\Delta^2}\eta R_0\Omega,
\label{eq:uxgas}
\end{equation}
\begin{equation}
u_y=-\frac{(1+\epsilon+\st^2)}{\Delta^2}\eta R_0\Omega,
\end{equation}
\begin{equation}
v_x=\frac{-2 \st}{\Delta^2}\eta R_0\Omega,
\label{eq:vxdust}
\end{equation}
\begin{equation}
v_y=-\frac{(1+\epsilon)}{\Delta^2}\eta R_0\Omega,
\label{eq:uydust}
\end{equation}
where $\Delta^2=\st^2+(1+\epsilon)^2$. For a fixed grain  size, these correspond to the traditional steady-state drift solutions for a dusty disc obtained by \citet{1986Icar...67..375N}.As mentionned above, including coagulation is not expected to  impact these equilibrium solutions, due to the presence of the sink term in the right-hand side of Eq. \ref{eq:mp}. 

\begin{table*} 
\caption{Summary of parameters adopted in our runs. }             
\label{table1}      
\centering                                  
\begin{tabular}{ccccccccccc}          
\hline\hline                        
  Run label  & Coagulation &${\it St_f}$& $\epsilon$  & $\alpha_{SS}$ & $\alpha_{g,x}$ &  $\alpha_{g,y}$ &  $\alpha_{g,z}$ & $\tau_{c,x}$ & $\tau_{c,y}$ &  $\tau_{c,z}$ \\ 
\hline                                  
E0v02S0v01 &no& $10^{-2}$ &  $0.02$   & -4.85e-07 &  1.12e-09& 3.57e-12 & 1.60e-12 & 0.14 & 1.53 & 0.34   \\
E0v02S0v01c &yes&  $10^{-2}$ &  $0.02$   & -9.45e-7 & 2.80e-7 & 3.06e-8 & 3.40e-7 & 0.12 & 0.18 & 0.22 \\
E0v02S0v1 &no& $10^{-1}$ &  $0.02$  & -4.80e-6 & 1.12e-9 & 2.73e-11 & 2.42e-13 & 0.10 & 1.37 & 0.42  \\
E0v02S0v1c &yes&  $10^{-1}$ &  $0.02$   & -8.55e-6 & 1.09e-7 & 5.11e-8 & 3.42e-6 & 0.09 & 0.17 & 0.12  \\
E0v02S1 &yes&   $1$ &  $0.02$   & -4.75e-5  & 3.04e-7 & 5.19e-7 & 2.06e-6& 0.10 & 0.11 & 0.3  \\
E0v02S1c &yes&  $1$ &  $0.02$   & -3.46e-5  & 8.89e-8 & 2.24e-7 & 7.92e-7& 0.09 & 0.10 & 0.12  \\
E0v2S0v1 &no& $10^{-1}$ &    $0.2$   & -5.5e-5 & 1.25e-8 & 1.26e-8 & 1.5e-8 & 0.16 & 0.19 & 0.17  \\
E0v2S0v1c &yes&  $10^{-1}$ &  $0.2$   & -1.06e-4 & 5.39e-6 & 1.31e-5 & 9.87e-6  & 0.09 & 0.09 & 0.11\\
E0v2S0v01 &no&$10^{-2}$  &  $0.2$   & -5.78e-6 & 1.17e-9 & 8.2e-11 & 6.33e-11 & 0.11 & 1.46 & 0.26 \\
E0v2S0v01c &yes&  $10^{-2}$ &  $0.2$  & -7.19e-6 & 2.15e-7 & 5.34e-7 & 5.63e-7 & 0.10 & 0.13 & 0.16 \\
E0v2S1 &no&   $1$ &  $0.2$   & -3.43e-4 & 5.61e-6 & 9.68e-6 & 6.93e-6 & 0.14 & 0.16 & 0.14 \\
E0v2S1c &yes& $1$ &  $0.2$  & -3.43e-4 & 5.29e-6 & 8.62e-6 & 6.41e-6 & 0.14 & 0.14 & 0.13 \\
E3S0v01 &no&   $10^{-2}$ &  $3$   & -3.43e-4 & 5.61e-6 & 9.68e-6 & 6.93e-6 & 0.17 & 0.24 & 0.09 \\
E3S0v01c &yes & $10^{-2}$ &  $3$   & -7.19e-6 & 2.15e-7 & 5.34e-7 & 5.63e-7 & 0.18 & 0.10 & 0.23 \\
E3S0v1 &no&  $10^{-1}$ &  $3$   & -1.1e-4 & 7.44e-06 &  7.89e-06 & 5.37e-06 & 0.08 & 0.24 & 0.13\\
E3S0v1c &yes& $10^{-1}$ &  $3$   &-1.1e-4 & 8.90e-06 & 8.18e-06 & 5.79e-06 & 0.08 & 0.26 & 0.13\\
E3S1 &no&  $1$ &  $3$   & -5.1e-6 & 9.52e-06 &  1.32e-05 & 4.37e-06 & 0.11 & 0.26 & 0.3\\
E3S1c &yes& $1$ &  $3$  &-5.1e-6& 9.58e-06& 1.32e-05 & 4.3e-06 & 0.08 & 0.24 & 0.13\\
\hline                                             
\end{tabular}
\label{table}
\end{table*}

\section{ Methodology} 
\subsection{Numerical method and model setup.}
We examine the nonlinear dynamical  evolution of the system by performing 2.5D (axisymmetric), unstratified simulations  in a shearing-box using the multifluid version of the publicly available code  FARGO3D \citep{2016ApJS..223...11B,2019ApJS..241...25B}. We considered single dust species runs in which the effect of coagulation was implemented in the aforementioned single-size approximation, albeit using a conservative version of Eq. \ref{eq:mp}:
\begin{equation}
\frac{\partial m_p \rho_d}{\partial t}+\nabla \cdot (m_p \rho_d \vdust)=\peff (\source-\sink) \rho_d
\label{eq:mpsinknum}
\end{equation}
Under this form, the previous equation can be integrated using the operator splitting technique employed by FARGO3D, with a coagulation term solved as a new substep within the source step. Tests of the implementation are presented in Appendix  \ref{sec:appA}. \\
With respect to Eq. \ref{eq:mp} , the previous expression includes an additional, sticking efficiency, term $\peff$ whose expression is similar to that  used by \citet{2022ApJ...937...21T}: \begin{equation}
\peff={\rm min}\left(1,-\frac{\ln(\st/\st_f)}{\ln 5}\right),
\label{eq:dustcoag}
\end{equation}
and which prevents the value for the Stokes number to increase above $\st_f$. This enables to i) take into account the effect of the fragmentation process in the simulations and ii) examine the evolution outcome as a function of parameter $\st_f$.

In our shearing-box setup, the cartesian coordinate system is oriented with $x$ along the (outward) radial direction and $z$ pointing along the (upward) vertical direction.  The domain considered is such that $x\in [-0.0125, 0.0125]$ and $z\in[-0.0025,0.0025]$, or equivalently $x\in(-0.25,0.25)H_0$ and $z\in(-0.05,0.05)H_0$. The appropriate size of the shearing-box was determined from convergence tests but typically,   a large radial  extent of the domain is needed to accommodate the most unstable wavelength expected from linear growth maps of the streaming instability as $\st \rightarrow \st_f$, whereas the small vertical extent of the domain is consistent with an  unstratified setup, focusing on the disc midplane. The domain is covered by $N_x \times N_z=2048 \times 512$ grid cells, with the radial resolution chosen as to capture the largest number of CI unstable modes, as these have growth rates increasing with radial wavenumber in absence of diffusive processes.

As we do not consider the vertical gravity in this work, boundaries are taken to be periodic in  the  z-direction, while we use standard shearing periodic boundaries in x.  
\subsection{ Parameters and initial conditions}
We adopt a unique value for the pressure gradient parameter $\eta=0.005$, corresponding to a typical pressure gradient in a protoplanetary disc. Three values of the initial  dust-to-gas mass ratio $\epsilon$ are considered, $\epsilon=0.02, 0.2, 3$,  which are representative of low ($\epsilon=0.02$), moderate ($\epsilon=0.2$), and high dust abundances ($\epsilon=3$).  For simulations in which dust coagulation is discarded, the Stokes number is kept constant with values: $\st=0.01, 0.1, 1$; whereas in those including coagulation the Stokes number is computed from Eq. \ref{eq:mpsinknum}, starting from an initial value $\st_i=10^{-3}$.
Details of parameters for each simulation are presented in Table \ref{table}. 

Regarding the initial conditions, radial and azimuthal components of dust and gas velocities are initialized according to Eqs. \ref{eq:uxgas}-\ref{eq:uydust}, while setting the vertical component to zero. In most runs,  the initial dust velocities are then seeded
with random noise of amplitude $10^{-2} c_s$. 
Because the effects of stratification are not considered, we use an initial gas density $\rho_0=1$.  We also use a system of units in which  $c_s=\Omega_k=H_g=1$. 
\subsection{Diagnostics}
One aim of this work is to assess the level of turbulence generated by the coagulation and streaming instabilities. To quantify the flux of angular momenum carried by turbulence, we first define the Shakura-Sunyaev stress parameter \citep{1973A&A....24..337S}: 

\begin{equation}
<\alphass>=\frac{<\rho_g u_x u_y>}{\rho_0 c_s^2},
\end{equation}

where the symbol $<>$ denotes averaging in the x-z plane. To obtain a reliable measure of $\alpha_{SS}$,  we further perform a time average of the latter quantity, and given by:
\begin{equation}
\overline{ <\alphass>}=\frac{1}{t_2-t_1}\int_{t_1}^{t_2} <\alphass>dt
\end{equation}
which we typically compute over a time interval of $\Delta t=t_2-t_1\sim 10 \omegam$ once saturation has been attained.  We also measure the velocity dispersion of the gas $\delta \bf u$ whose {\it i}th component is given by: 
\begin{equation}
\delta u_i=\sqrt{<u_i^2>-<u_i>^2}, 
\label{eq:deltau}
\end{equation}
and the autocorrelation time $t_{c,i}$ of the velocity fluctuations, which is obtained by evaluating each component of the autocorrelation function given by \citep{2018ApJ...868...27Y}:
\begin{equation}
{\cal R}_i(t)=\int [u_i(\tau)-\overline{u_i}][u_i(t+\tau)-\overline{u_i}] d\tau
\label{eq:tcor}
\end{equation}
In the previous expression,  $\overline u$ is the mean velocity, whose components $u_i$ are estimated by considering their time average in the interval $[t_1,t_2]$:
\begin{equation}
\overline{u_i}=\frac{1}{t_2-t_1}\int_{t_1}^{t_2} u_i(\tau) d\tau
\label{eq:tcor2}
\end{equation}
Following \citet{2022ApJ...937...55H}, we set $t_2-t_1=5P$ and record the data with a cadence of $0.01P$ (Yang et al. 2018) to compute Eq. \ref{eq:tcor}. We then determine $t_{c,i}$ by fitting an exponential function to the autocorrelation function \citep{2006A&A...452..751F}. 

From Eqs. \ref{eq:deltau} and \ref{eq:tcor}, we can subsequently compute  the gas bulk diffusion coefficient $D_{g,i}$ in the {\it i}th direction and which is given by:

\begin{equation}
D_{g,i}=\overline{\delta u_i}^2t_{c,i}
\end{equation}

In the following, we will rather make  use of the dimensionless bulk gas diffusion coefficient, defined by: 
\begin{equation}
\alpha_{g,i}=\frac{D_{g,i}}{c_s H_g}=\left(\frac{\overline{\delta u_i}}{c_s}\right)^2\tau_{c,i}
\end{equation}

where $\tau_{c,i}$ is the dimensionless equivalent of the  correlation time:

\begin{equation}
\tau_{c,i}=t_{c,i}\Omega_0^{-1}
\end{equation}

\begin{figure*}
\centering
\includegraphics[width=0.33\textwidth]{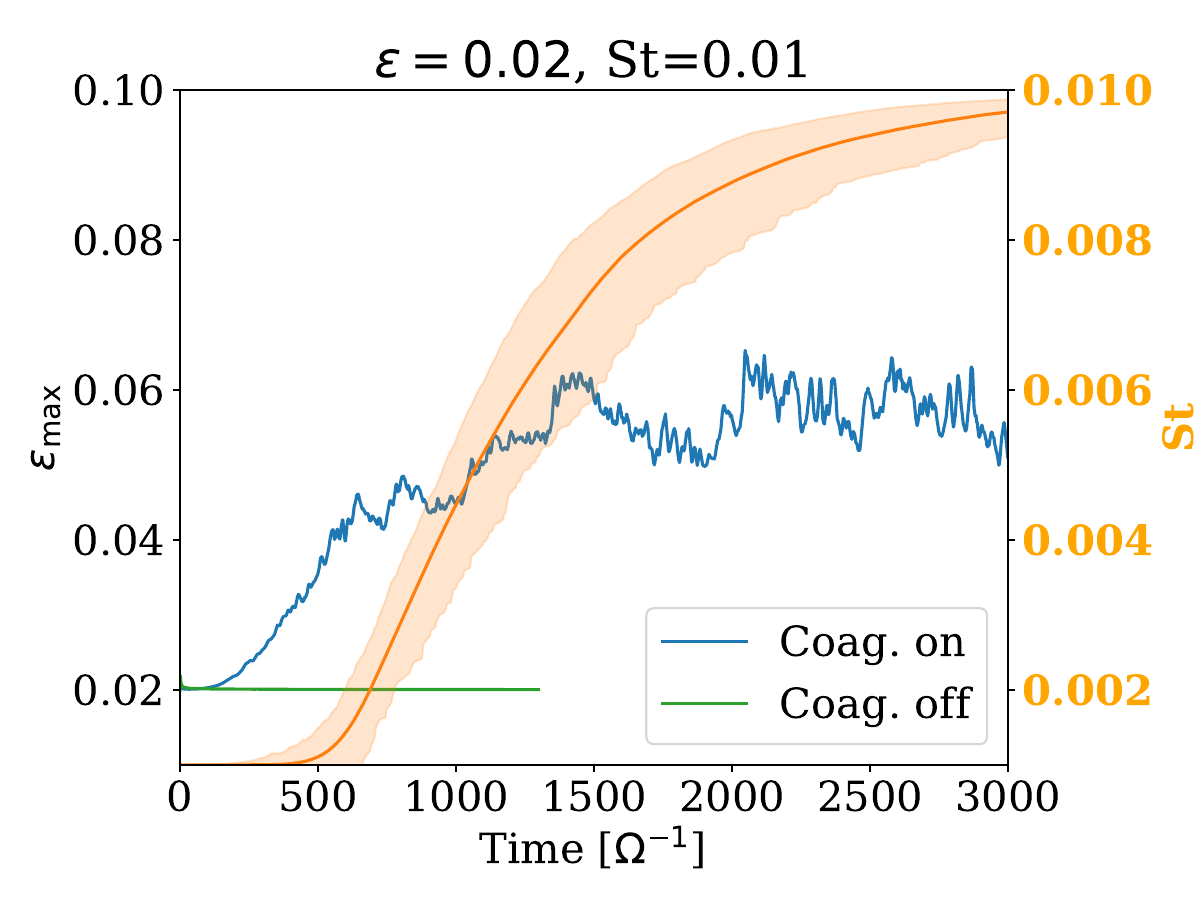}
\includegraphics[width=0.33\textwidth]{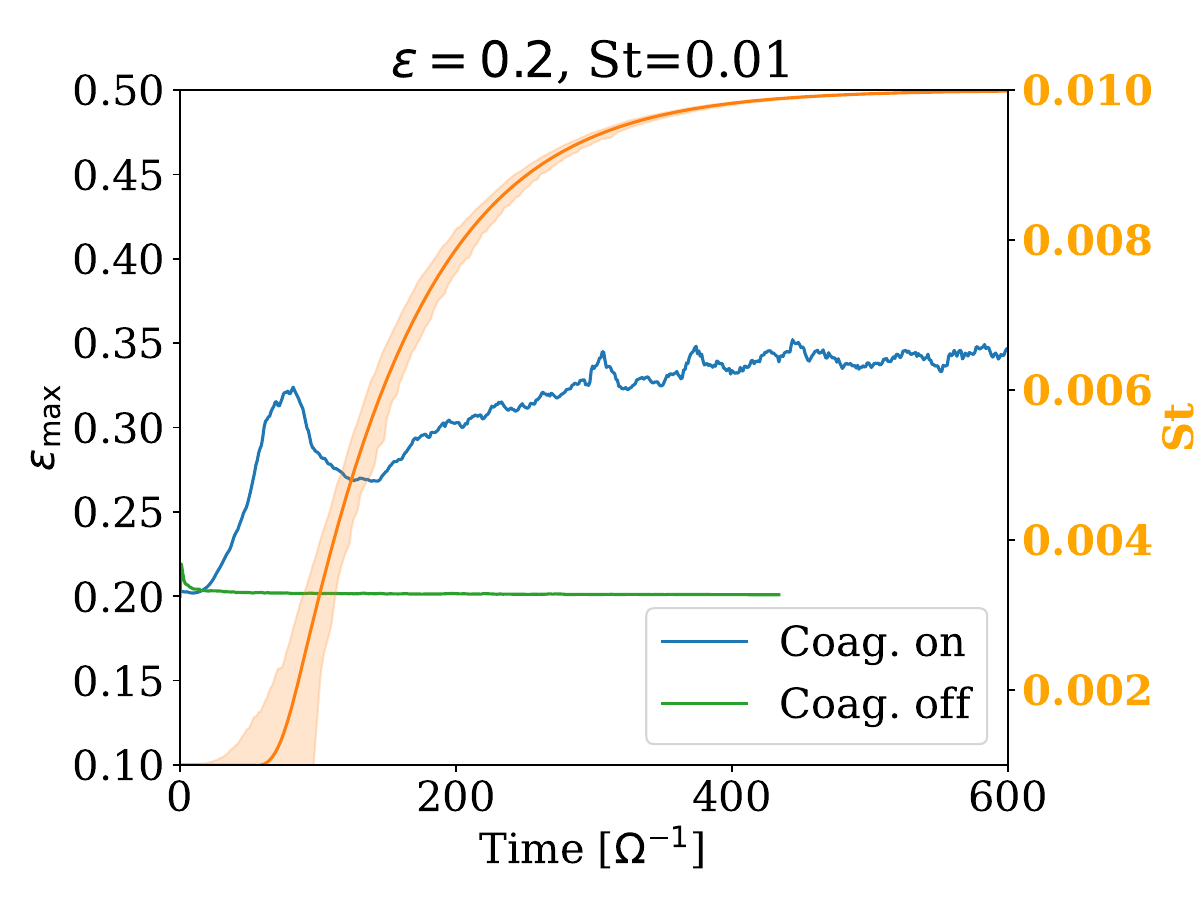}
\includegraphics[width=0.33\textwidth]{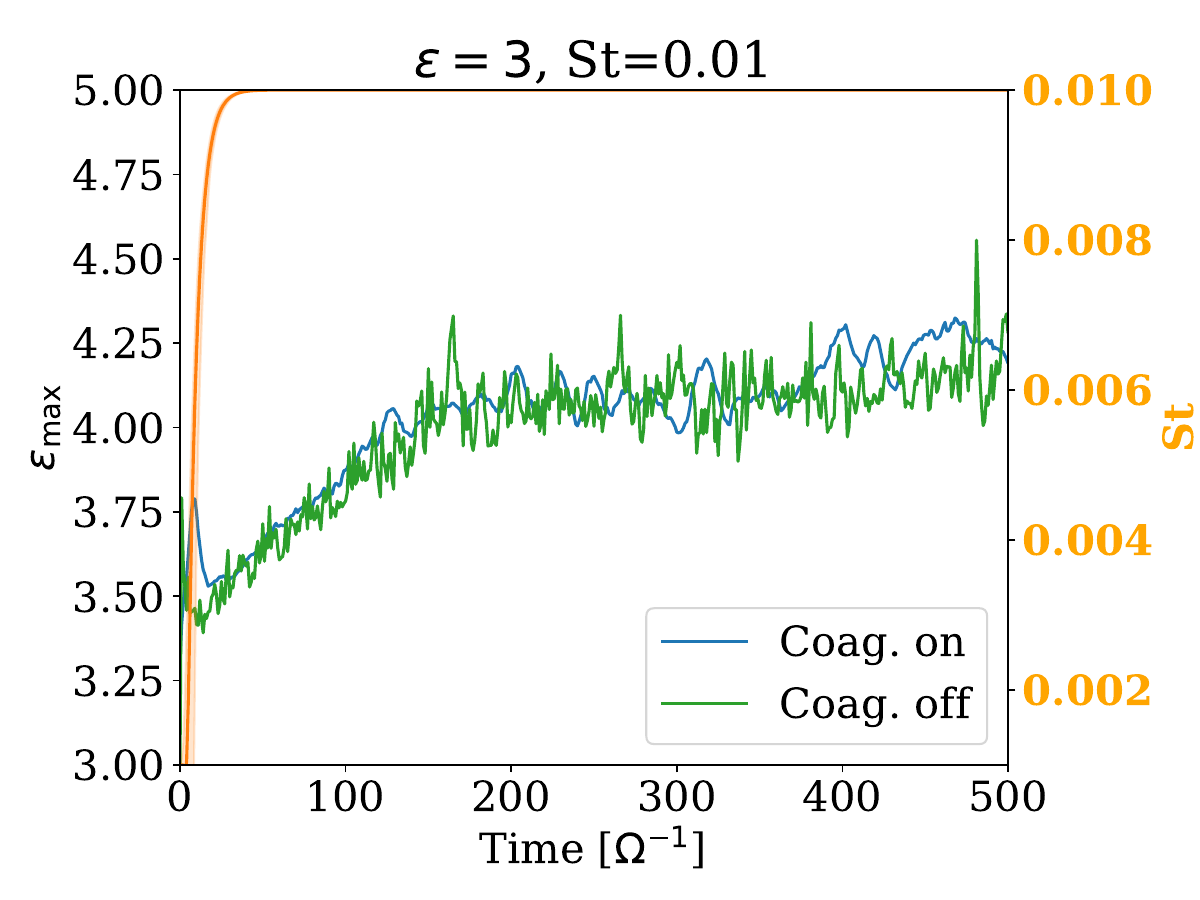}
\includegraphics[width=0.33\textwidth]{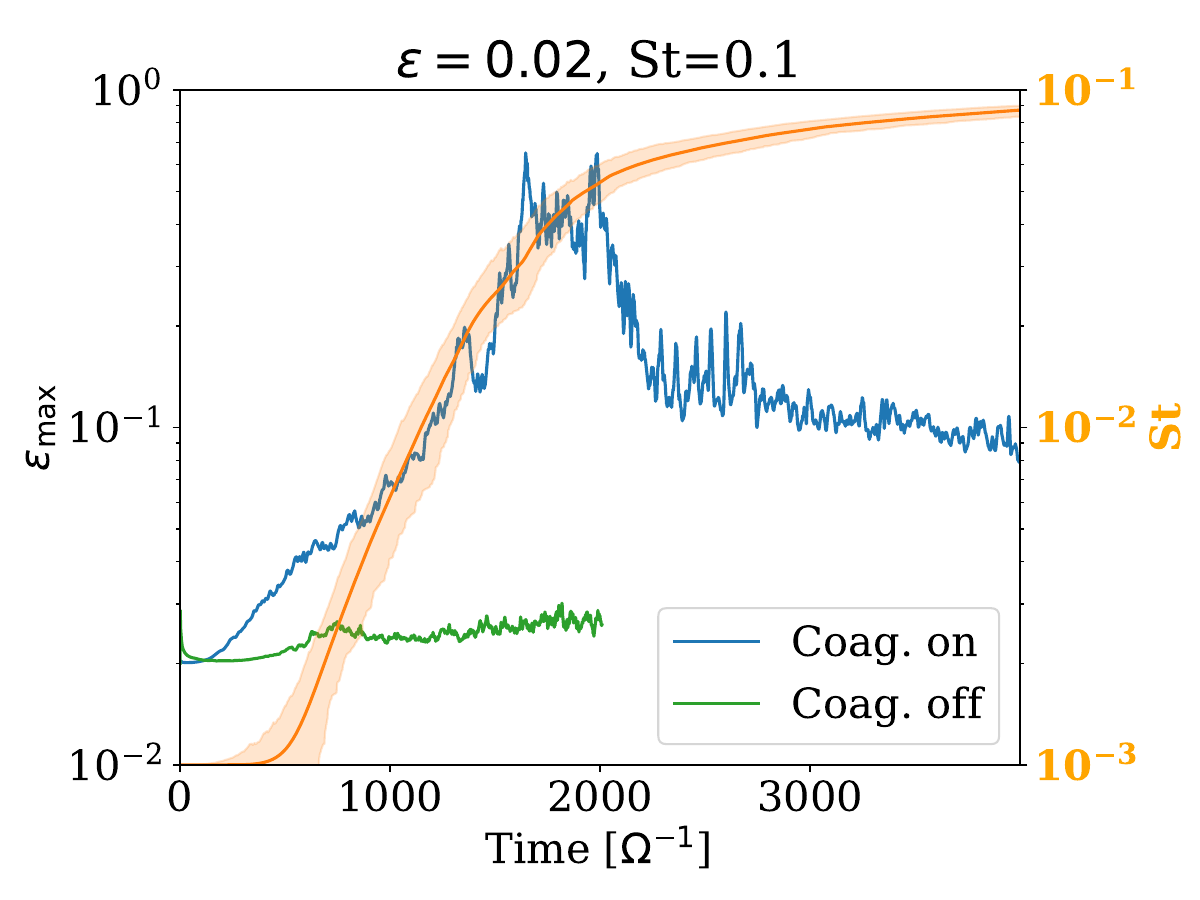}
\includegraphics[width=0.33\textwidth]{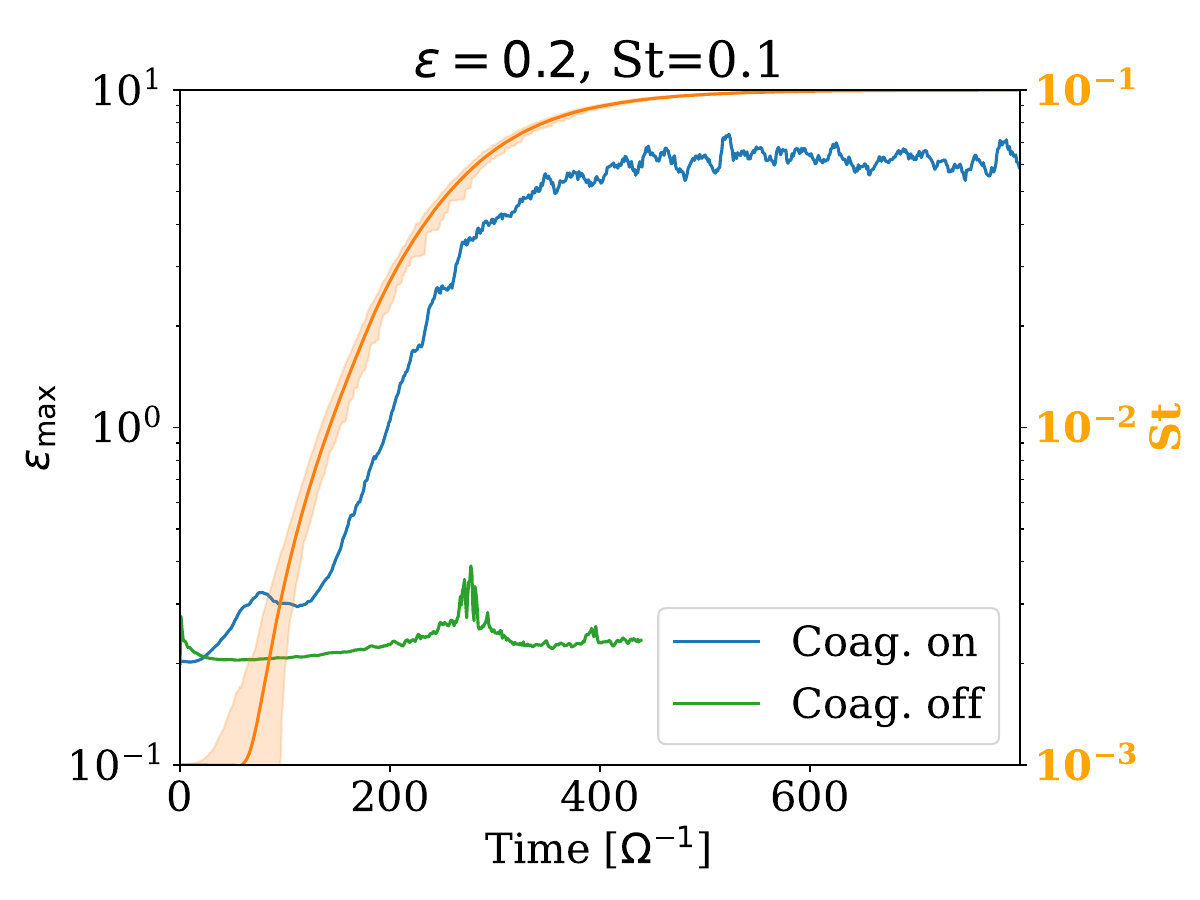}
\includegraphics[width=0.33\textwidth]{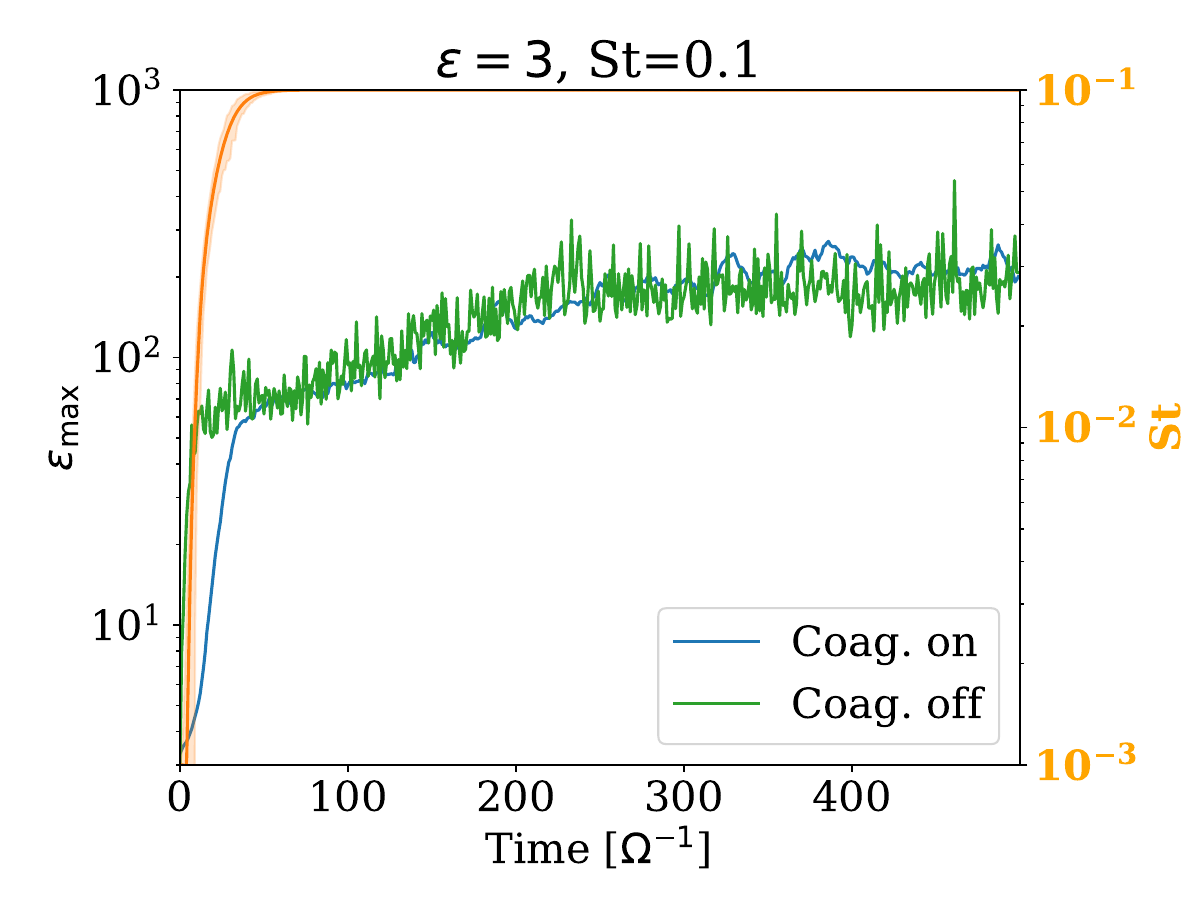}
\includegraphics[width=0.33\textwidth]{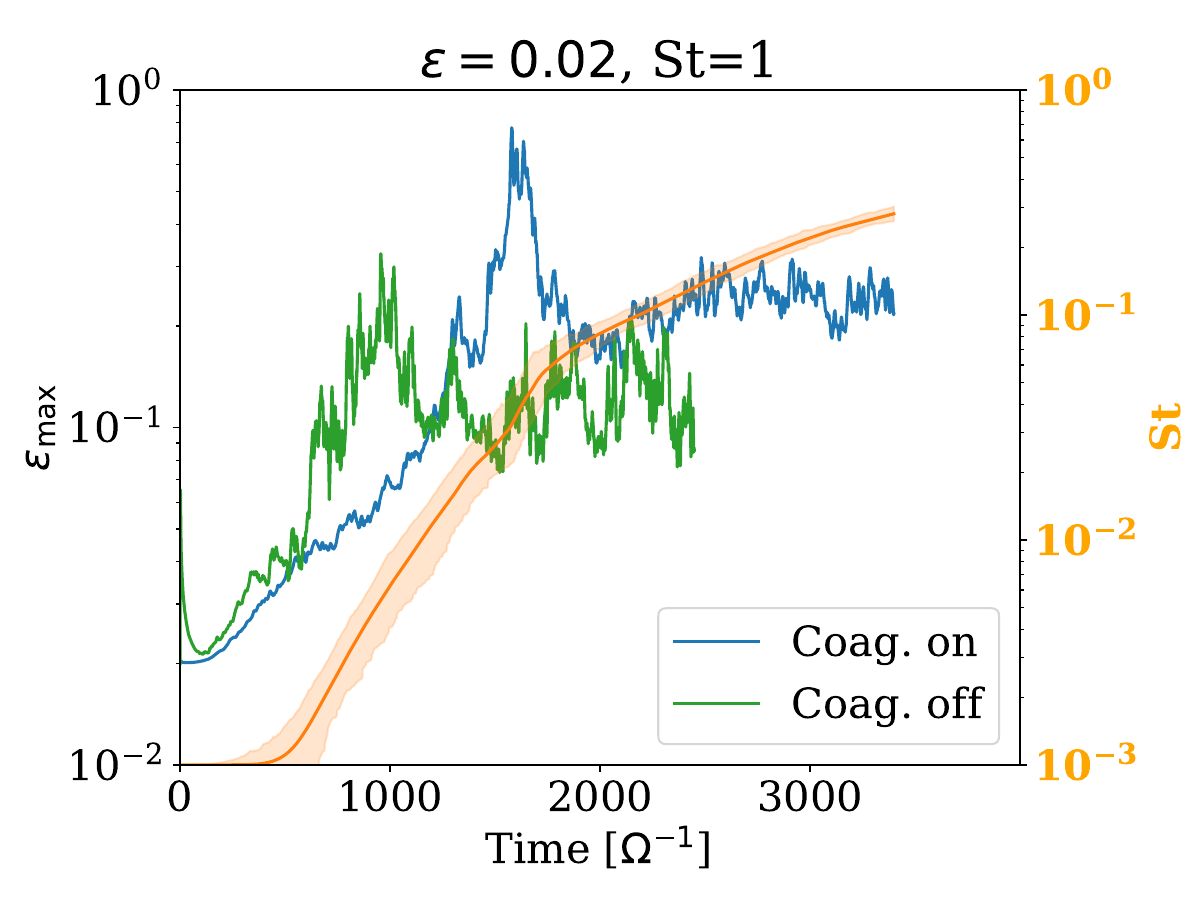}
\includegraphics[width=0.33\textwidth]{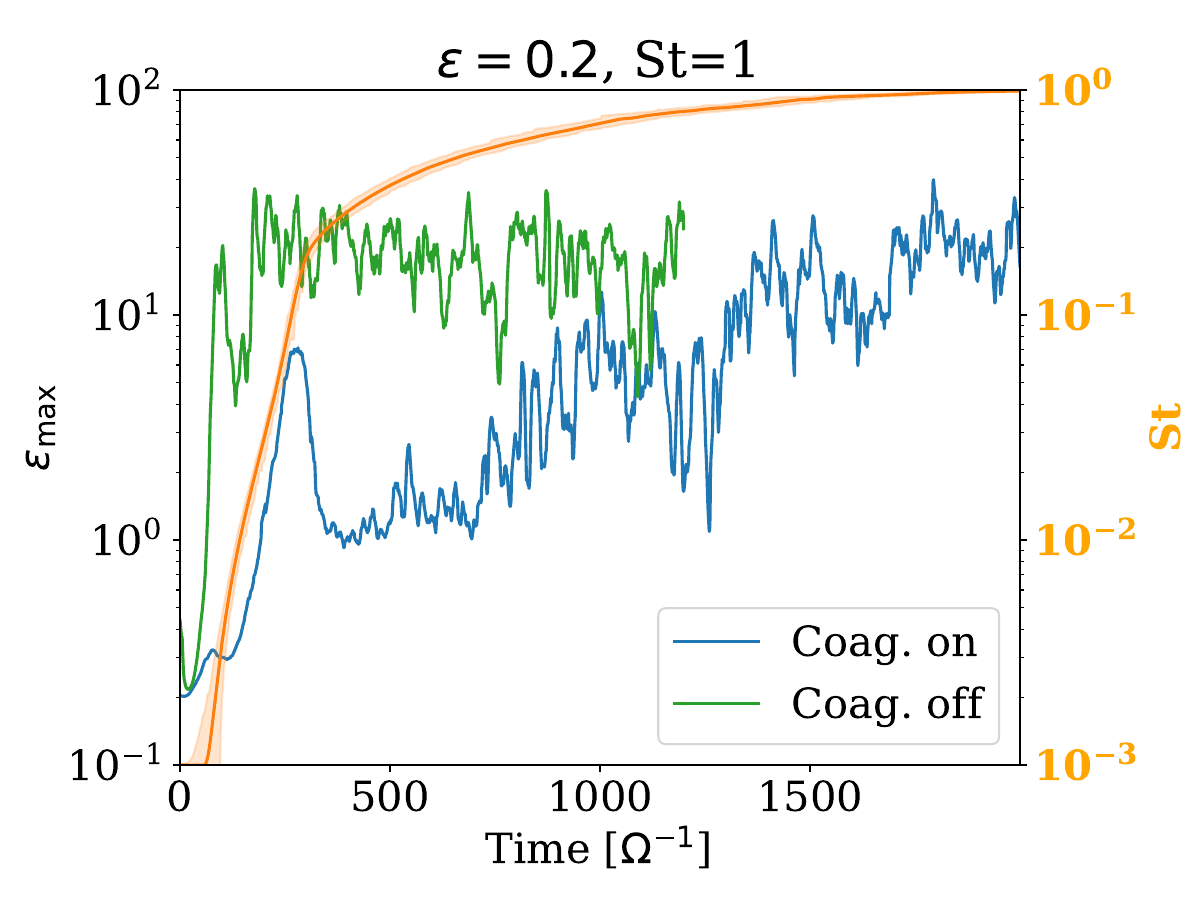}
\includegraphics[width=0.33\textwidth]{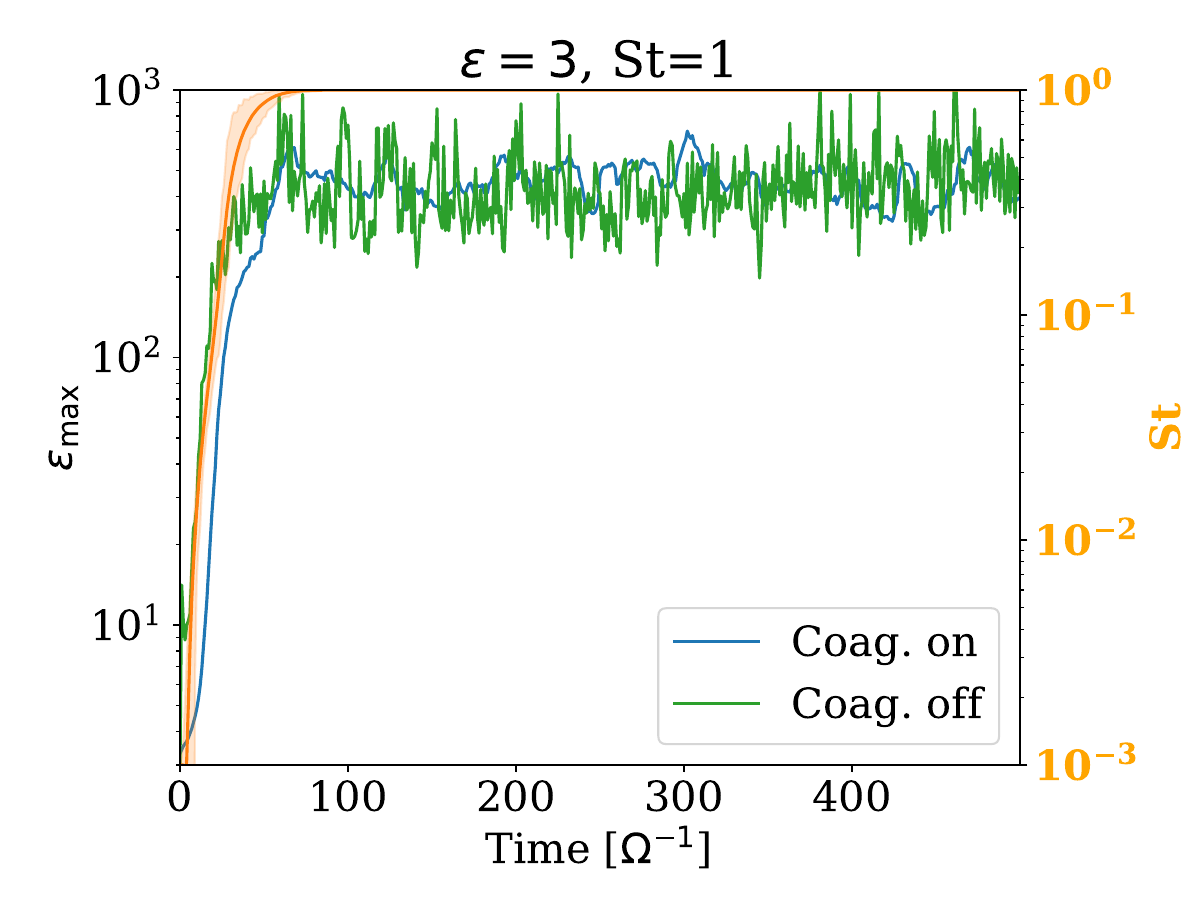}
\caption{Time evolution of  the maximum dust-to-gas ratio $\epsilon_{max}$ and Stokes number for all simulations listed in Table \ref{table}. The left, middle, right columns correspond to $\epsilon=0.02, 0.2, 3$ respectively while the upper, middle, and lower panels correspond to $\st=0.01, 0.1, 1$ respectively. In each panel, we show the evolution of $\epsilon_{max}$ for the case with (blue)  and without (green) coagulation. For the evolution of $\st$, the light colour traces the minimum and maximum values whereas the dark colour traces the box averaged value. }
\label{fig:dustmax}
\end{figure*}

\section{Results}
In this section, we discuss the impact of dust coagulation on the non-linear evolution of the SI by providing  a systematic comparison between  pure SI simulations and runs that include dust coagulation.  Our comparison rely on models in which the maximum Stokes number $\st_f$ coincides with the value for  $\st$ set in standard SI simulations. For simplicity, we will therefore simply denote $\st_f$ as $\st$ when presenting the results of the simulations.  

In Fig. \ref{fig:dustmax} we present an overview of the results where we  show the time evolution of the maximum dust-to-gas ratio $\epsilon_{max}$ and Stokes number $\st$ for each model we considered.  From this figure, it is immediately obvious that in dust rich discs, the impact of coagulation is relatively weak as simulations that include this effect reach overdensities that are very similar to those obtained in standard SI simulations. 

In dust-poor discs with $\epsilon < 0.2$, however, including coagulation has a stronger effect, with overdensities that can be a factor of $5-30$ higher than without coagulation for tightly coupled particles. In this regime, we find that the coagulation instability (hereafter CI; \citet{2021ApJ...923...34T}) plays an important role in the evolution, as it is found to  dominate over other processes for $\epsilon=0.02$, or even interact with the SI for $\epsilon=0.2$. Below, we  describe in more details these two different modes of evolution found  for  Stokes numbers $\st <0.1$. We then discuss the case of marginally coupled particles with $\st\approx 1$.

\begin{figure*}
\centering
\includegraphics[width=0.49\textwidth]{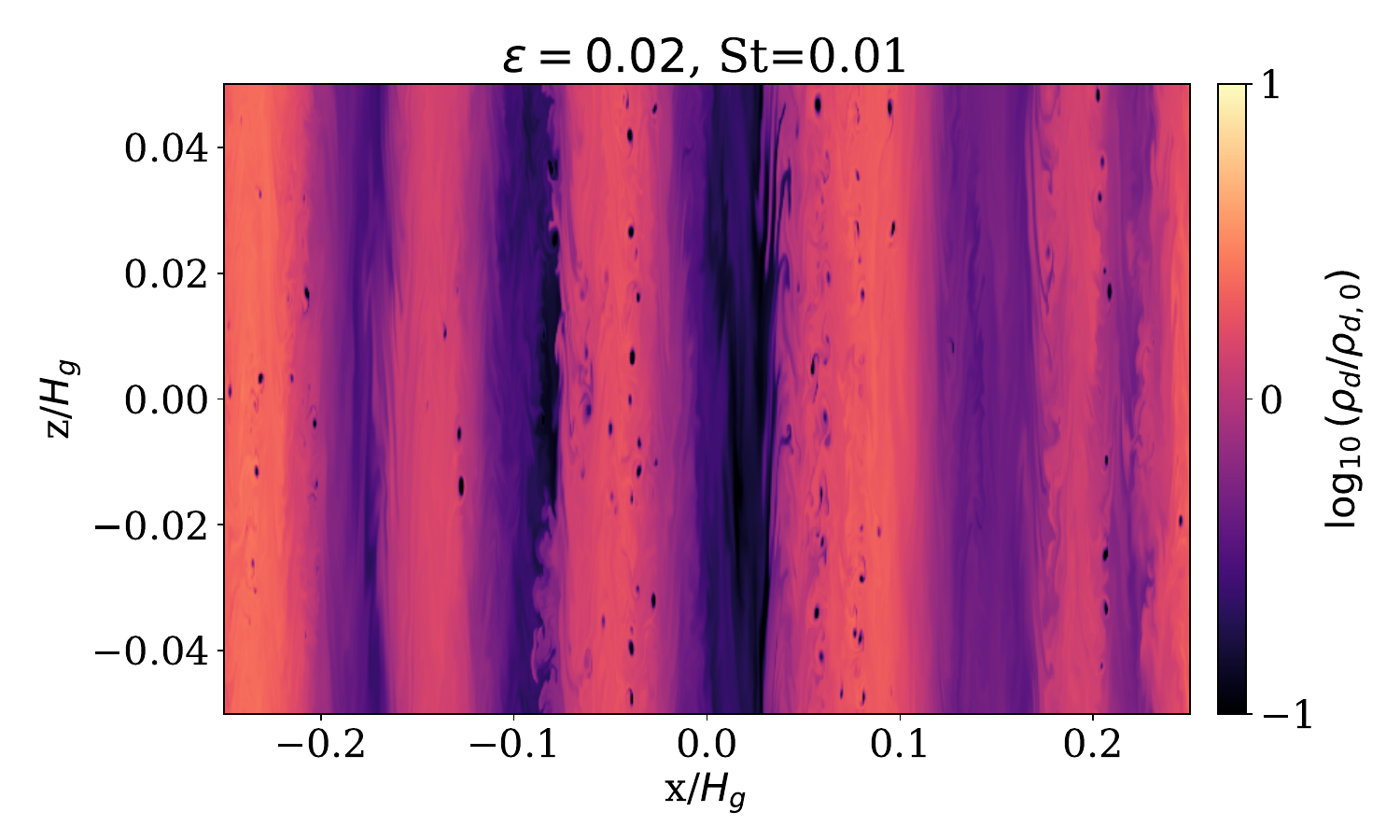}
\includegraphics[width=0.49\textwidth]{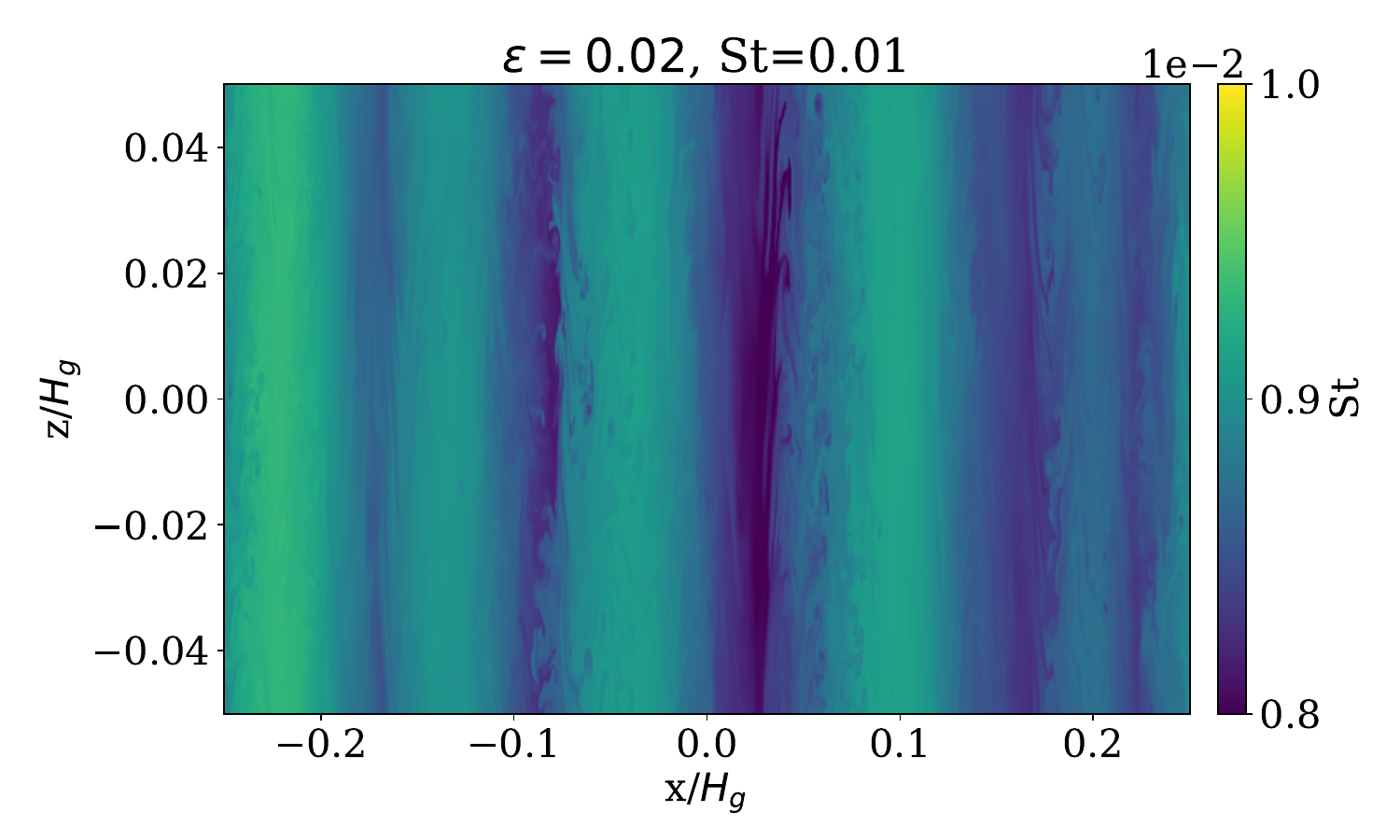}
\includegraphics[width=0.49\textwidth]{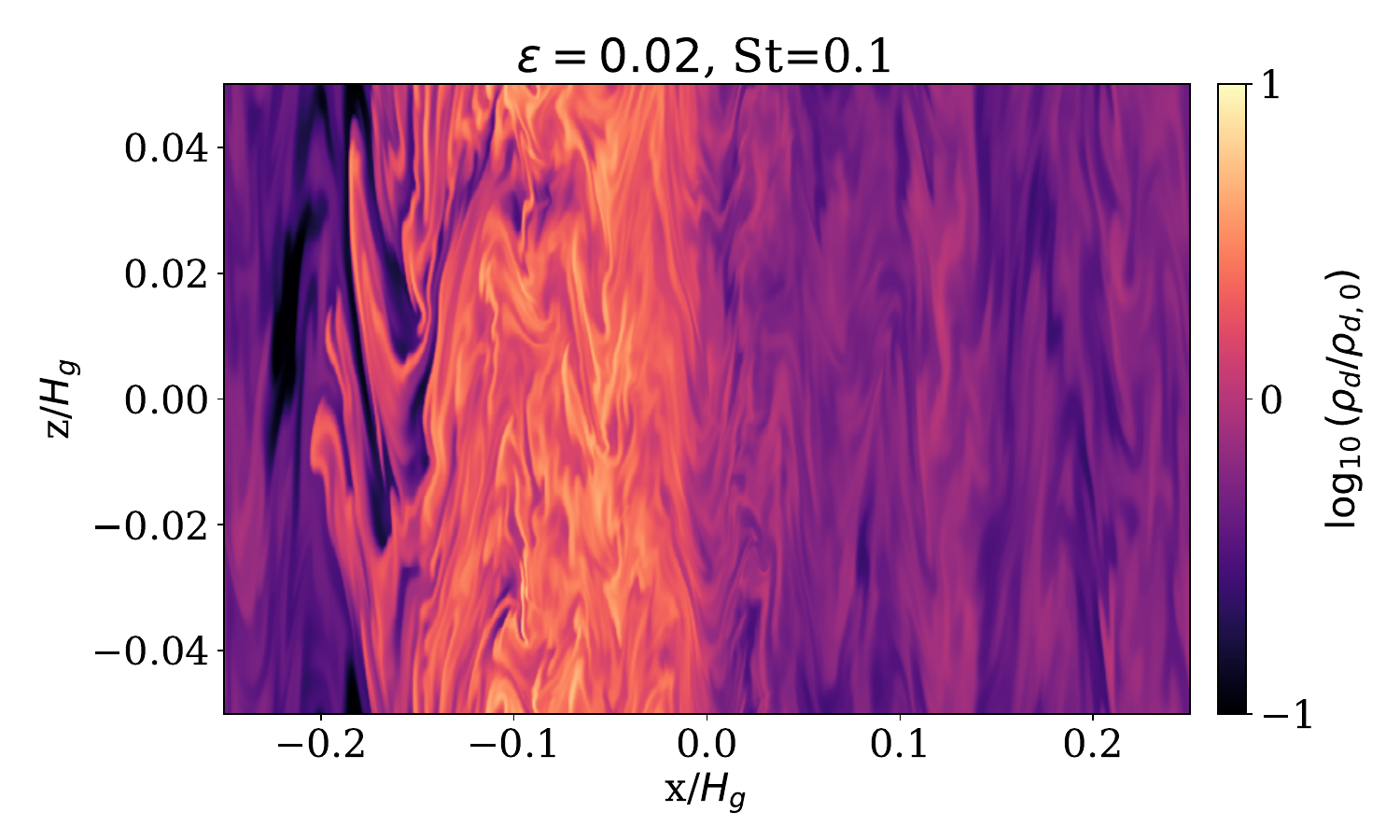}
\includegraphics[width=0.49\textwidth]{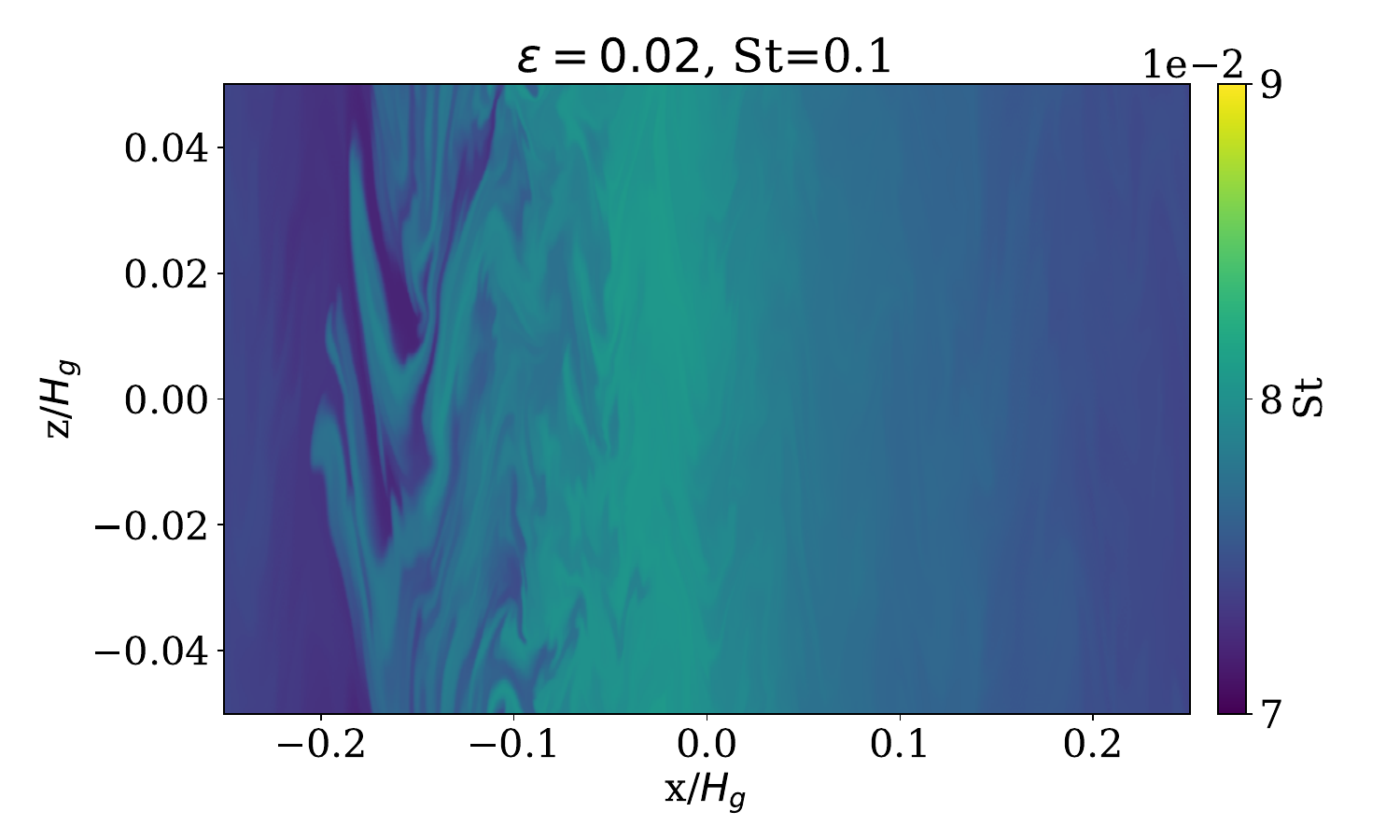}
\caption{In the case $\epsilon=0.02$, snapshots of the dust density (left) and Stokes number (right) for model E0v02S0v01c with $\st=0.01$ (left), and model E0v02S0v1c with $\st=0.1$ (right).  Both models include the effect of coagulation and result in the growth of the coagulation instability. }
\label{fig:2deps0v02}
\end{figure*}

\begin{figure}
\centering
\includegraphics[width=0.49\columnwidth]{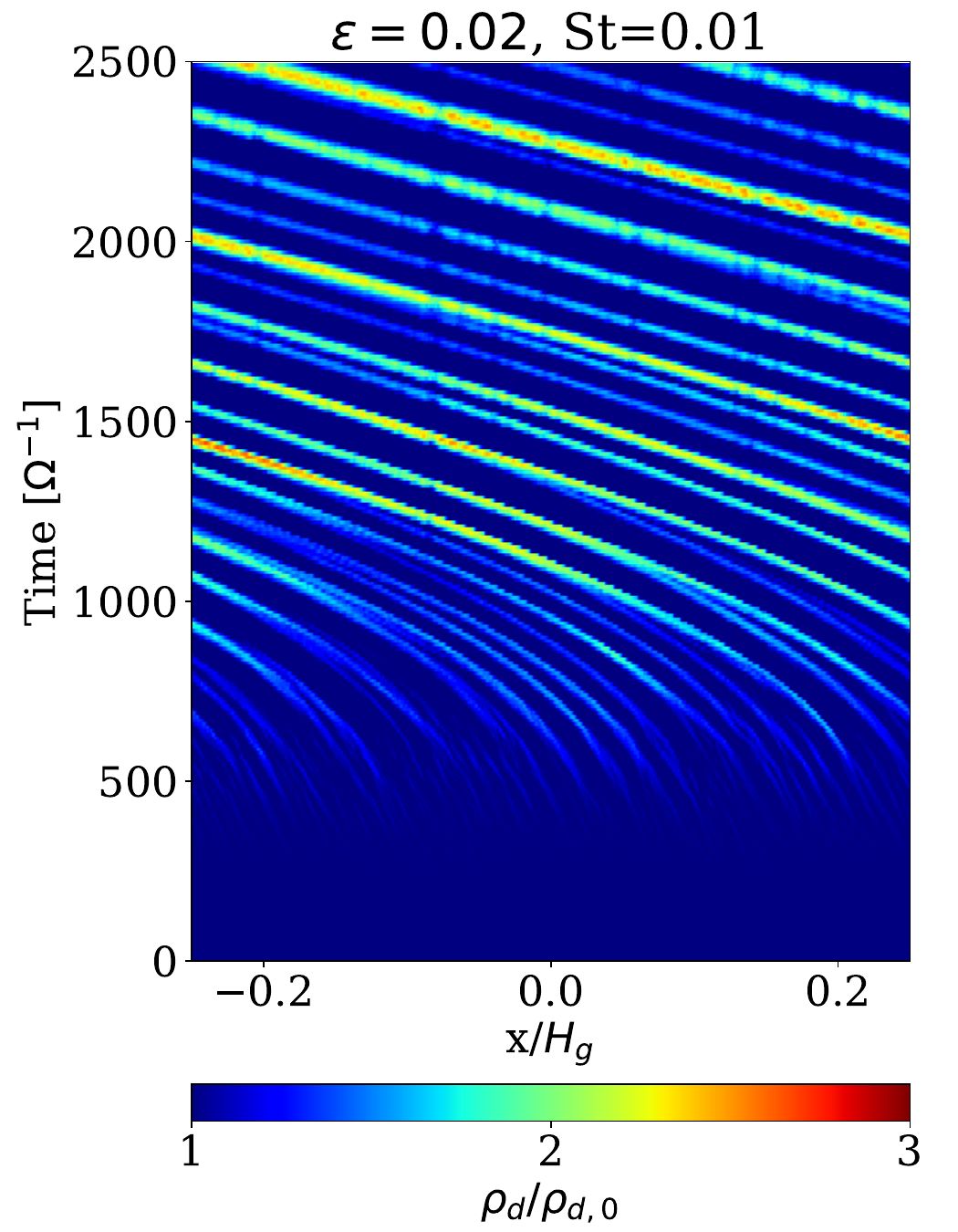}
\includegraphics[width=0.49\columnwidth]{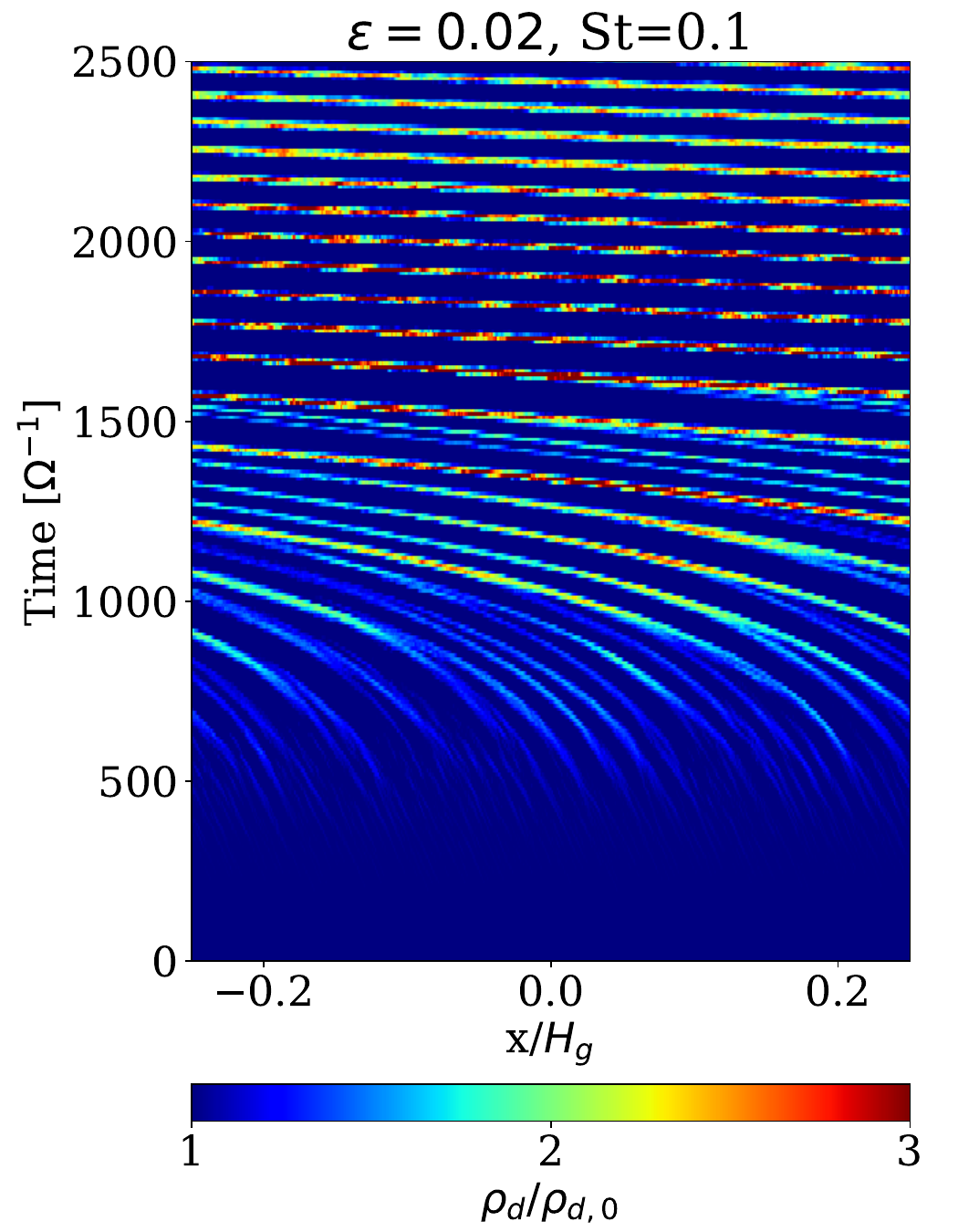}
\caption{In the case $\epsilon=0.02$, time evolution of the vertically averaged dust density for model E0v02S0v01c with $\st=0.01$ (left), and model E0v02S0v1c with $\st=0.1$ (right).  }
\label{fig:spacet}
\end{figure}

\begin{figure}
\centering
\includegraphics[width=\columnwidth]{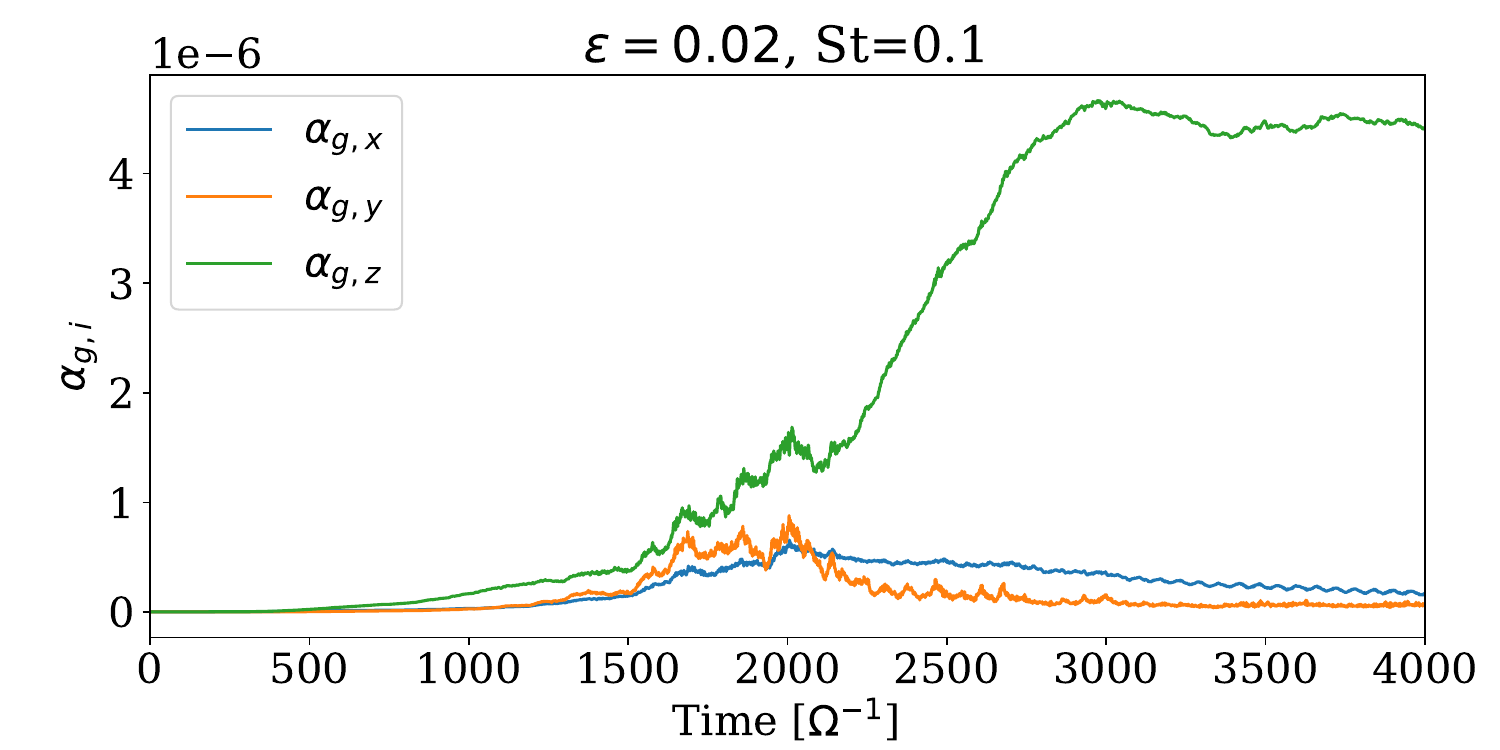}
\includegraphics[width=0.49\columnwidth]{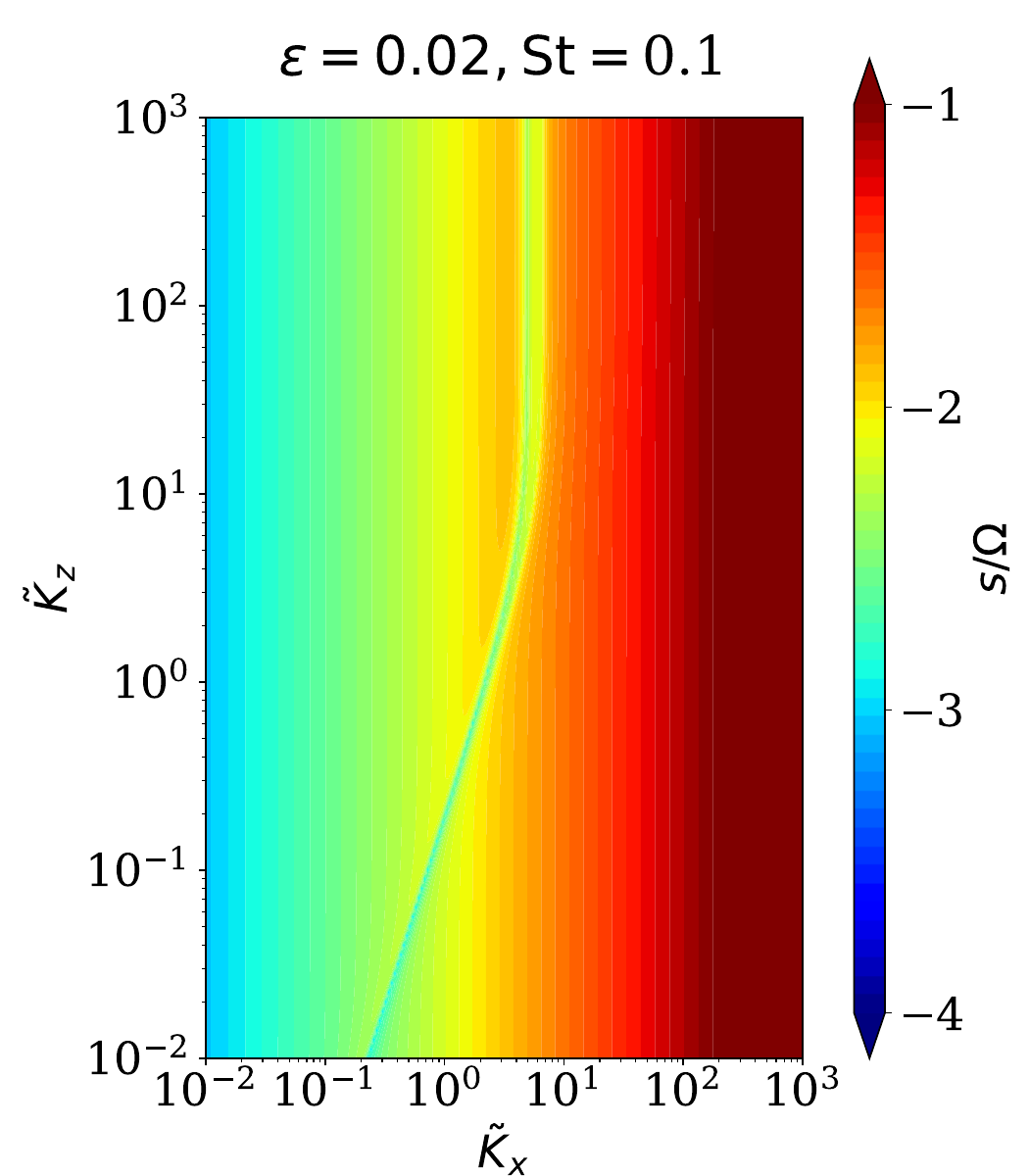}
\includegraphics[width=0.49\columnwidth]{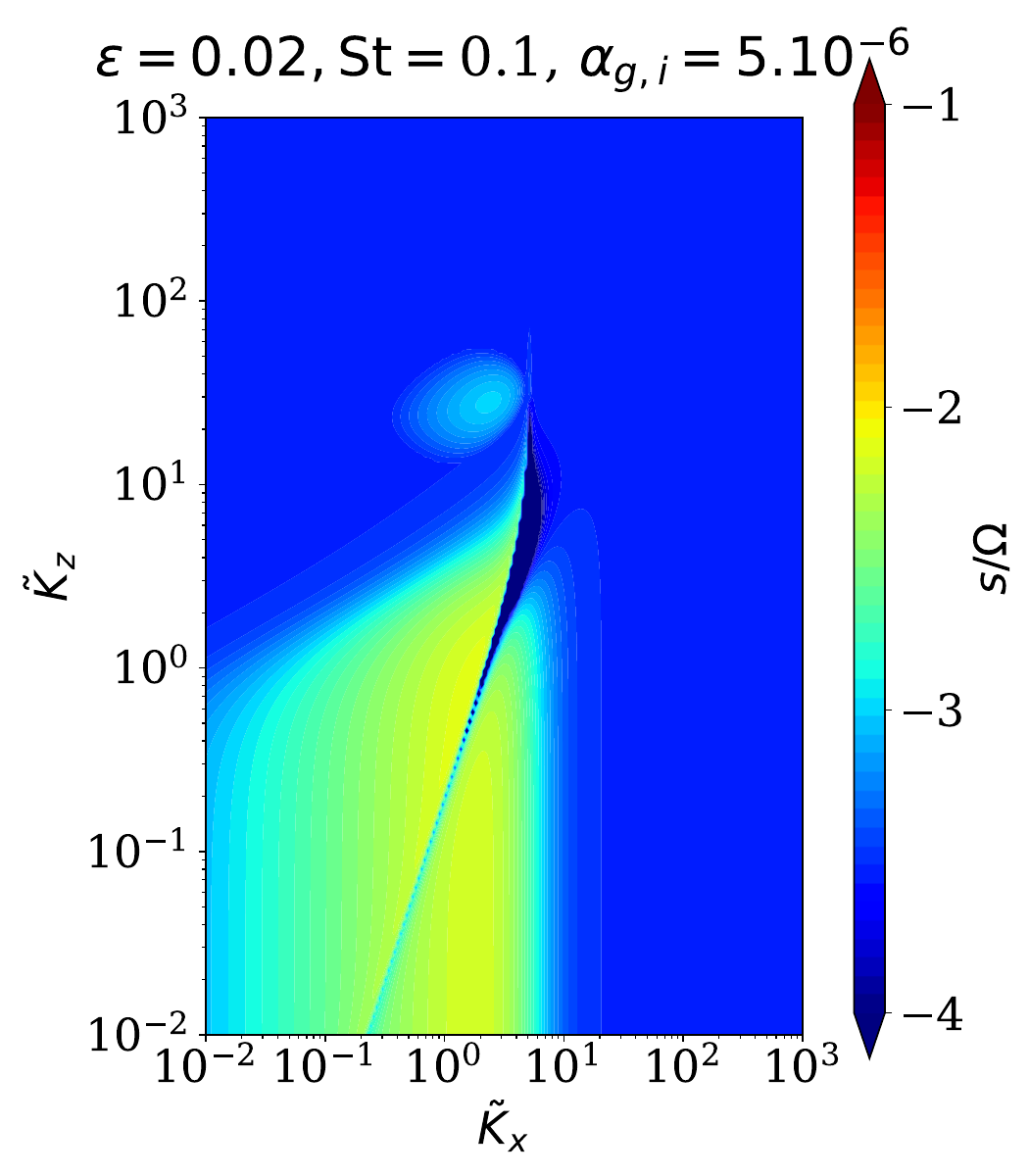}
\caption{For $\epsilon=0.2$ and $\st=0.01$ (model E0v02S0v01c) , time evolution of the dimensionless diffusion coefficients $\alpha_{g,i}$ in each direction (top). The lower panel shows expected linear growth rates for the same model in the inviscid case (left) and for  $\alpha_{g,i}=5\times 10^{-6}$ (right).  These have been obtained from linear theory (see App. \ref{sec:appA}) using $D=\nu=\alpha_{g,i}c_s H_g$. Here, $\tilde K_{x,z}=k_{x,z} \eta R_0$ are the dimensionless wavenumbers.  }
\label{fig:e0v02s0v1diff}
\end{figure}

\subsection{Tightly coupled particles}
\subsubsection{$\epsilon=0.02$: Evolution mediated by CI}
\label{sec:411}

The CI is by essence a dust-driven  instability which involves two ingredients:   i) the dust-density density dependence of the collisional coagulation rate and ii) the dependence of $\st$ with the particle size.  When these are mixed, a dust surface density perturbation $\delta \Sigma_d$ leads to a radial variation of $\st$, resulting in  a traffic jam which tends to enhance $\delta \Sigma_d$. Compared to other dust-driven instabilities, the CI can develop over a few orbital periods even for small dust-to-gas ratios $\epsilon\sim 10^{-3}$. More precisely, the growth rate of the CI is typically given by \citep{2021ApJ...923...34T}: 
\begin{equation}
\sigma=\sqrt{\frac{1}{2}\frac{\epsilon}{3 t_0}k |<v_{x}>|}\Omega^{-1}
\end{equation}
where $k$ is the radial wavenumber,  and $v_x$ is given by Eq. \ref{eq:vxdust}. $t_0=4/3\sqrt{2\st_i/\alpha C \pi}$ is the mass growth timescale of a dust particle and can be obtained from Eq. \ref{eq:mp}.  For $\st_i=10^{-3}$ and $\epsilon=0.02$, and setting $kH=4\times 10^3$ which corresponds to the maximum wavenumber  that is resolved by $\approx 6$ grid cells, we get $\sigma\approx 2.10^{-3}\Omega^{-1}$. The time evolution of the maximum dust-to-gas ratio and Stokes number is shown, for $\epsilon=0.02$ and $\st=0.01$ (Model E0v02S0v01c),  in the upper left panel of Fig. \ref{fig:dustmax}.  Both quantities exhibit an initial rapid exponential growh phase and then reach non-linear saturation at $t\sim 1500\Omega^{-1}$. We note in passing that in the context of the CI, unstable modes are essentially incompressible such that the evolution of $\epsilon_{max}$ reflects the evolution of the maximum of the perturbed dust density $\delta \rho_d$.   From this plot,   we estimate a linear growth rate of $\sigma\sim 1.7\times 10^{-3}\Omega^{-1}$ during the initial linear growth phase, which is in good agreement with our previous analytical estimation. This clearly indicates that the CI is responsible for the initial  increase in $\delta \rho_d$ and $\st $ that  are observed, although disentangling the effect of the CI from the evolution of the averaged Stokes number is more difficult. For such small values of $\epsilon$ and $\st$, we also note  that the effect of the SI is  expected to be negligible, and this is confirmed by inspecting the evolution for the  case where coagulation is switched off (green line), and for which $\epsilon_{max}$ remains close to its initial value. 

Fig. \ref{fig:2deps0v02} shows snapshots of the dust density and Stokes number at saturation, for this model and in the case with $\epsilon=0.02$ and $\st=0.1$ (Model E0v02S0v01c) for which a similar mode of evolution is found. Since the CI is originally a one-dimensional instability with no dependence in the vertical direction,  it is not surprising to observe the formation of vertically extended filaments, whose number tends to decrease as $\st$ is increased. For $\st=0.1$, we indeed ultimately obtain a single filament with $\epsilon_{max} \approx 0.1$ whereas for $\st=0.01$,   five filaments remain in the system at the end of the simulation with consequently smaller $\epsilon_{max}\approx 0.006$. We find that these filaments are themselves formed through the merging of thinner ones. This is illustrated in Fig. \ref{fig:spacet} where we display the vertically-averaged dust density as a function of time and radial position. From this figure, it is clear that filaments grow not only via merging events but also by simply collecting dust as they drift inward.  Both effects then combine to make the amplitude of  the perturbed dust density start level off and saturate at later times. Another mechanism that may contribute to saturation is the level of  turbulence generated by the CI itself. This is suggested by looking back to Fig. \ref{fig:2deps0v02} where it is evident that turbulence operates in the disc, especially for $\st=0.1$. For this case, the time evolution of the dimensionless bulk gas diffusion coefficients $\alpha_{g,i}$ is shown in the upper panel of Fig. \ref{fig:e0v02s0v1diff}. Not surprisingly, this reveals anisotropy of the CI with an enhanced dust diffusion coefficient $\alpha_{g,z}$ of  ${\cal O}(10^{-6})$ in the vertical direction, which is larger by two orders of magnitude compared to $\alpha_{g,y}$. A similar trend is also found for $\st=0.01$ (see Table \ref{table}), with  $\alpha_{g,z}$ of ${\cal O}(10^{-7})$ and $\alpha_{g,y}$ of ${\cal O}(10^{-8})$ in that case. The bottom panel of Fig. \ref{fig:e0v02s0v1diff}  compares linear growth rates for $\alpha_{g,i}=0$ (left) and for $\alpha_{g,i}=3\times 10^{-6}$ (right), and which have been obtained from the linearized equations presented in Appendix \ref{sec:appA}  using $\nu=D=\alpha_{g,i}c_sH_g$. We see that the turbulent diffusion induced by the CI itself can indeed significantly cut off the most unstable modes corresponding to  $\alpha_{g,i}=0$. Although not shown here, we find a similar effect  for $\st=0.01$, which confirms that the CI-induced dust diffusion can also contribute to the saturation of the instability. 

\begin{figure}
\centering
\includegraphics[width=\columnwidth]{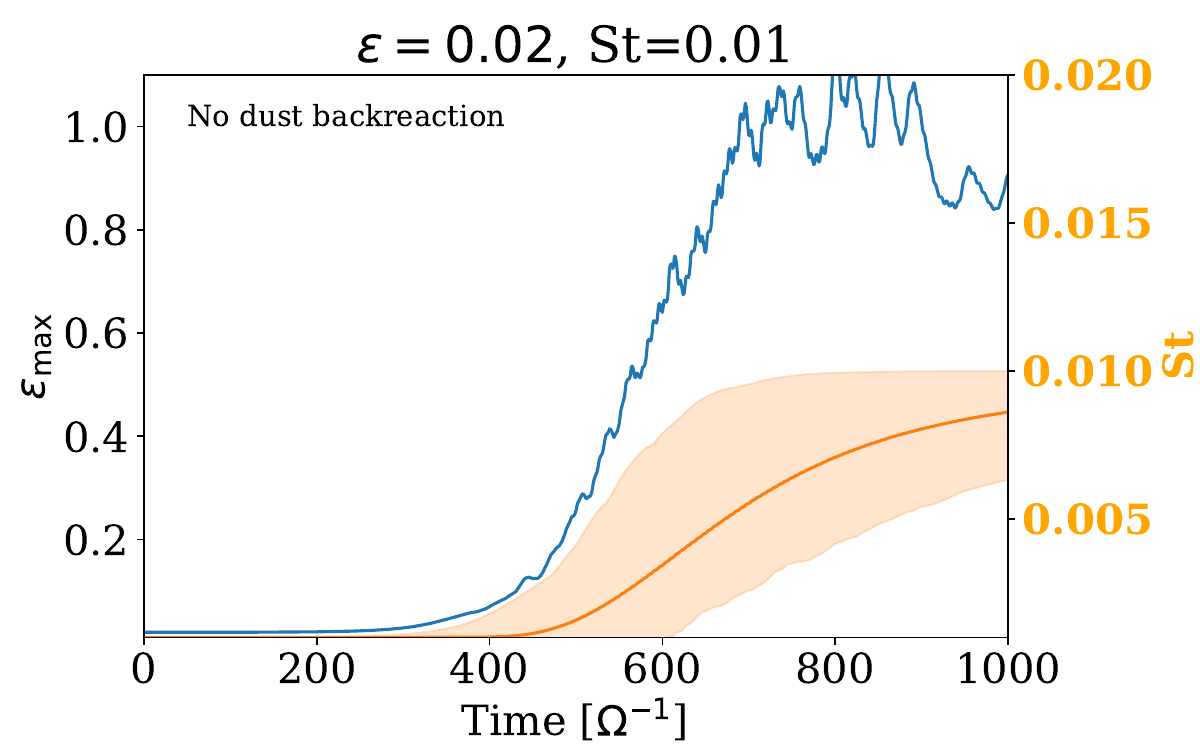}
\includegraphics[width=\columnwidth]{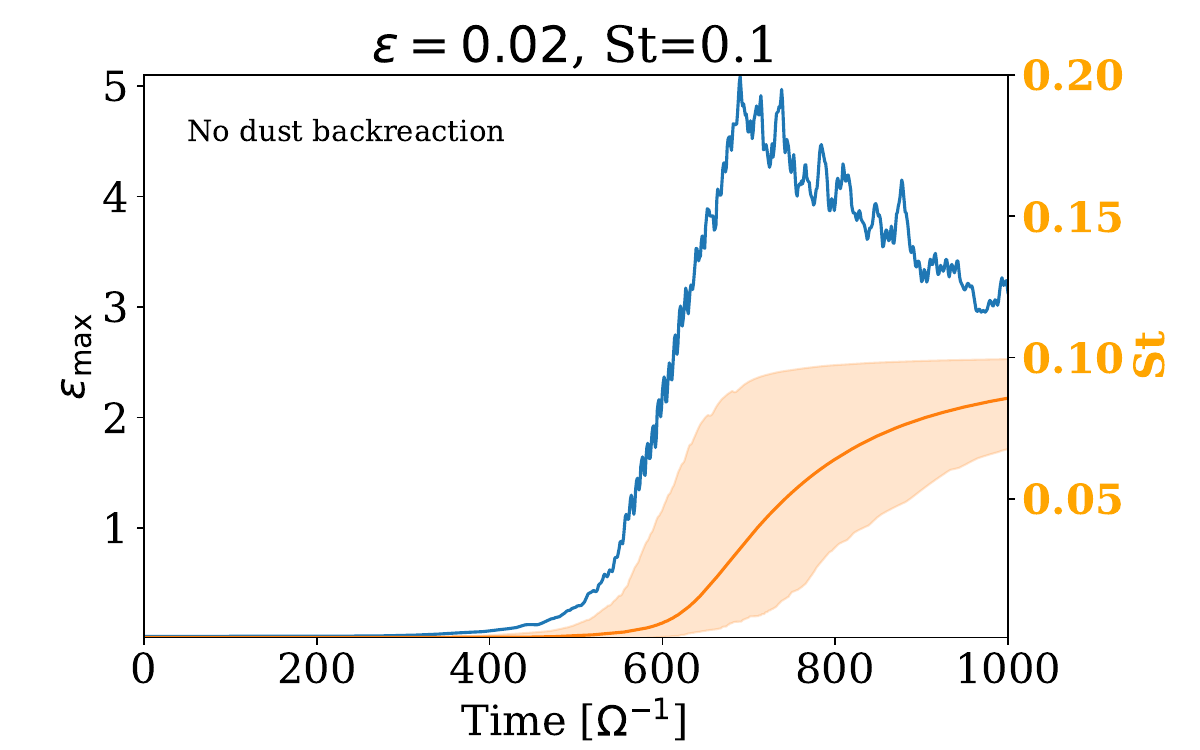}
\caption{For $\epsilon=0.2$, time evolution of  the maximum dust-to-gas ratio $\epsilon_{max}$ and Stokes number for $\st=0.01$ (top) and $\st=0.1$ (bottom), in the case where the effect of the dust backreaction onto the gas is discarded.}
\label{fig:nofeedback}
\end{figure}

In summary, we find that for tightly coupled particles, the CI can operate in dust-poor discs and lead to the formation of dust rings with $\epsilon_{max}$ of ${\cal O}(0.1)$, which is, however, not enough to subsequently trigger the SI within these rings. This is because the dust backreaction onto the gas tends to signicantly regulate the maximum dust concentration that can be achieved through the CI \citep{2022ApJ...940..152T}. This is illustrated  in Fig. \ref{fig:nofeedback}  where we show the time evolution of the maximum dust-to-gas ratio and Stokes number for the same models that include dust coagulation, but in the case where the effect of the dust backreaction onto the gas is discarded. We find a signicantly higher $\epsilon_{max}$ in that case, consistently with the results of \citet{2022ApJ...940..152T} (see their Figure 3.).  

\subsubsection{$\epsilon=0.2$: Interplay between CI /SI and turbulence}
\label{sec:e0v2}

\begin{figure}
\centering
\includegraphics[width=\columnwidth]{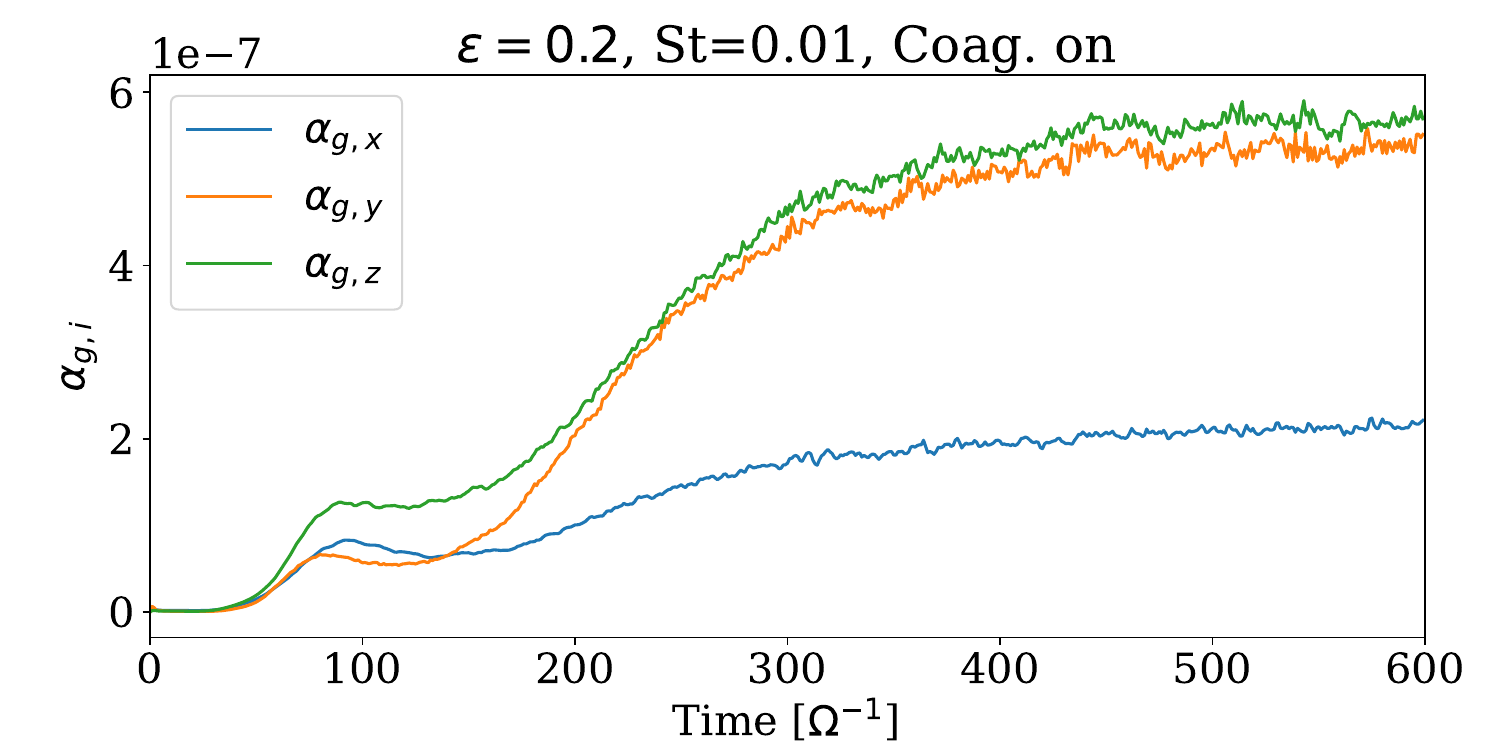}
\includegraphics[width=\columnwidth]{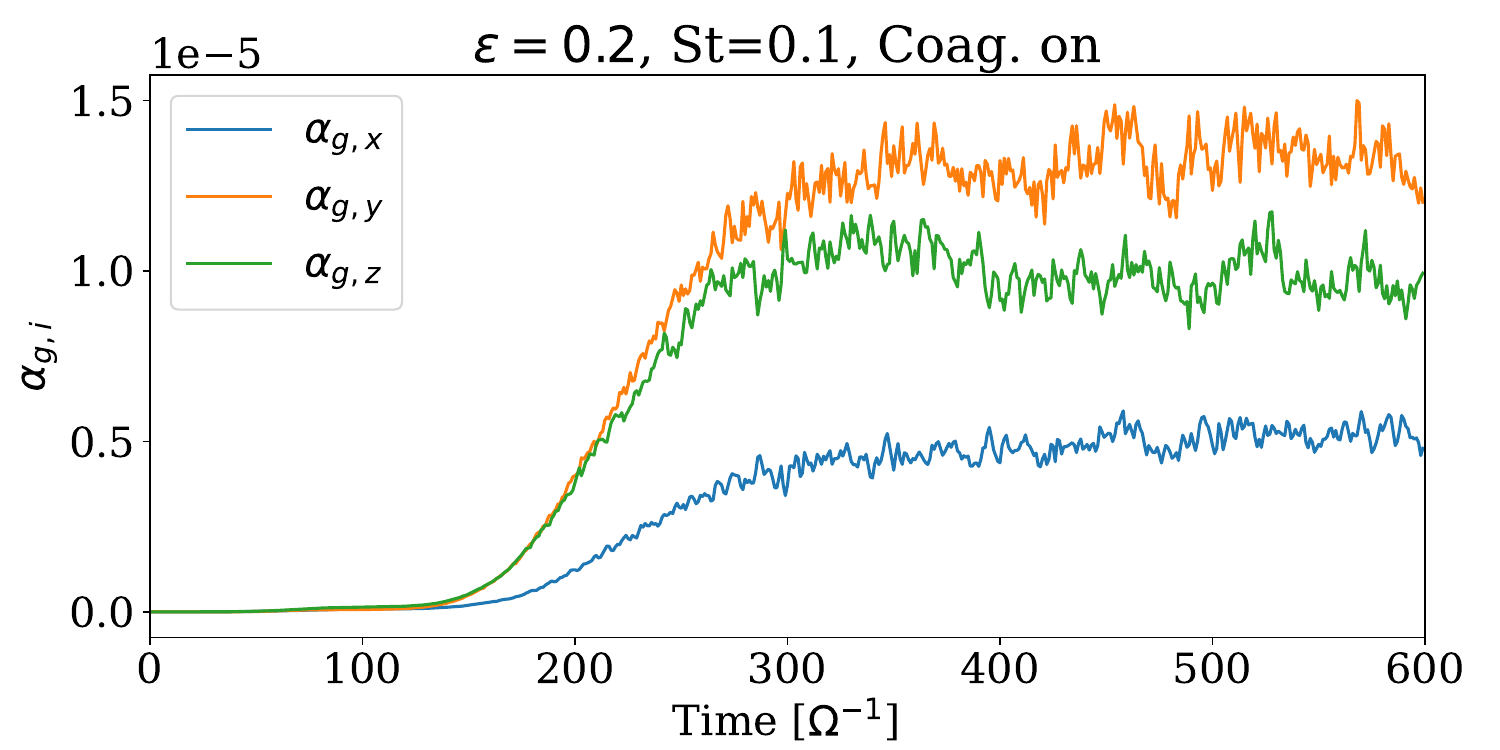}
\caption{For $\epsilon=0.2$, time evolution of the dimensionless diffusion coefficients $\alpha_{g,i}$ in each direction for model E0v2S0v01c with $\st=0.01$ (top) and model E0v2S0v1c with $\st=0.1$ (bottom).}
\label{fig:alphagi_eps0v2}
\end{figure}

\begin{figure}
\centering
\includegraphics[width=\columnwidth]{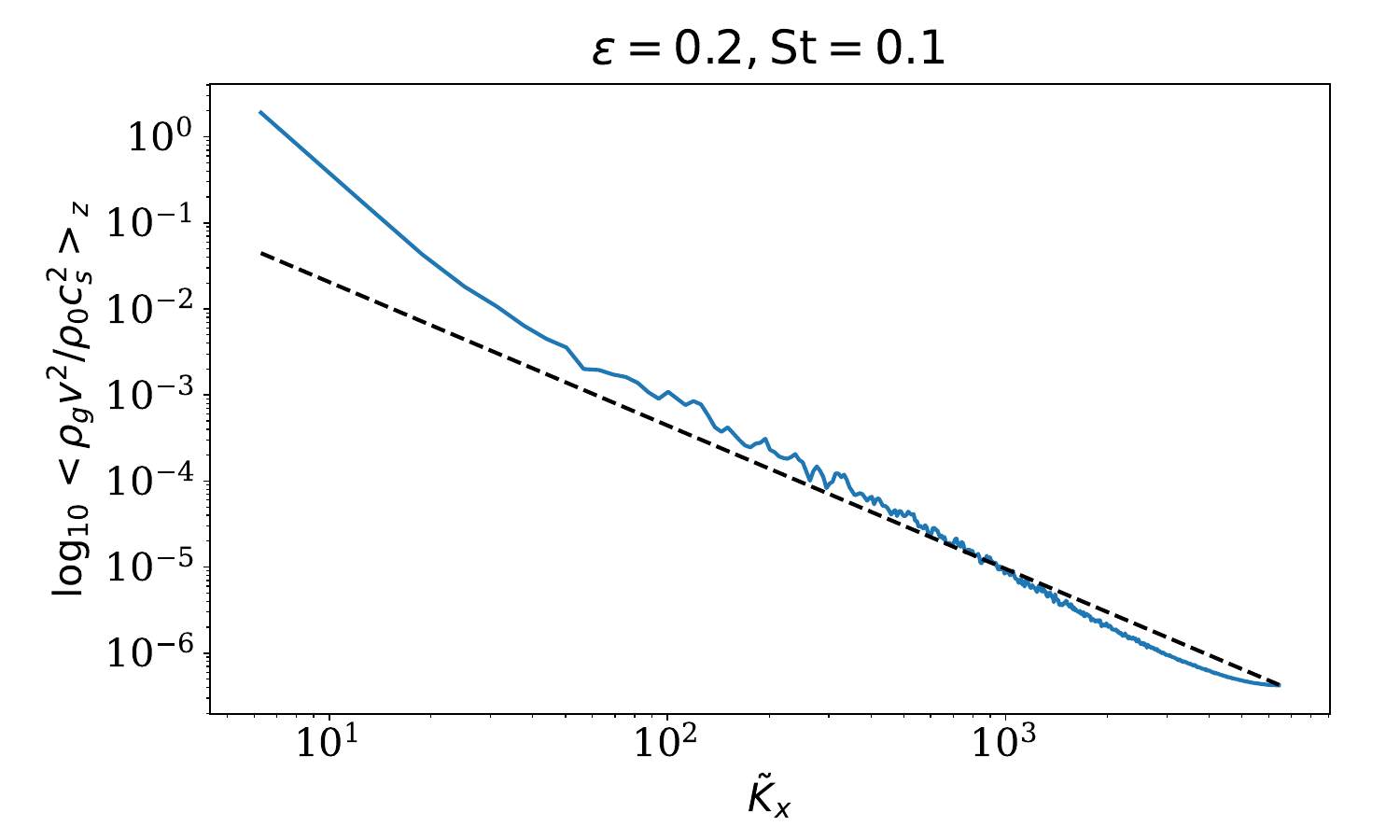}
\caption{Spectrum of the vertically averaged gas kinetic energy density for $\epsilon=0.2$ and $\st=0.1$ (run E0v2S0v1c). Here, $\tilde K_{x}=k_{x} \eta R_0$ is the dimensionless radial wavenumber.}
\label{fig:spectrum}
\end{figure}

\begin{figure*}
\centering
\includegraphics[width=0.7\textwidth]{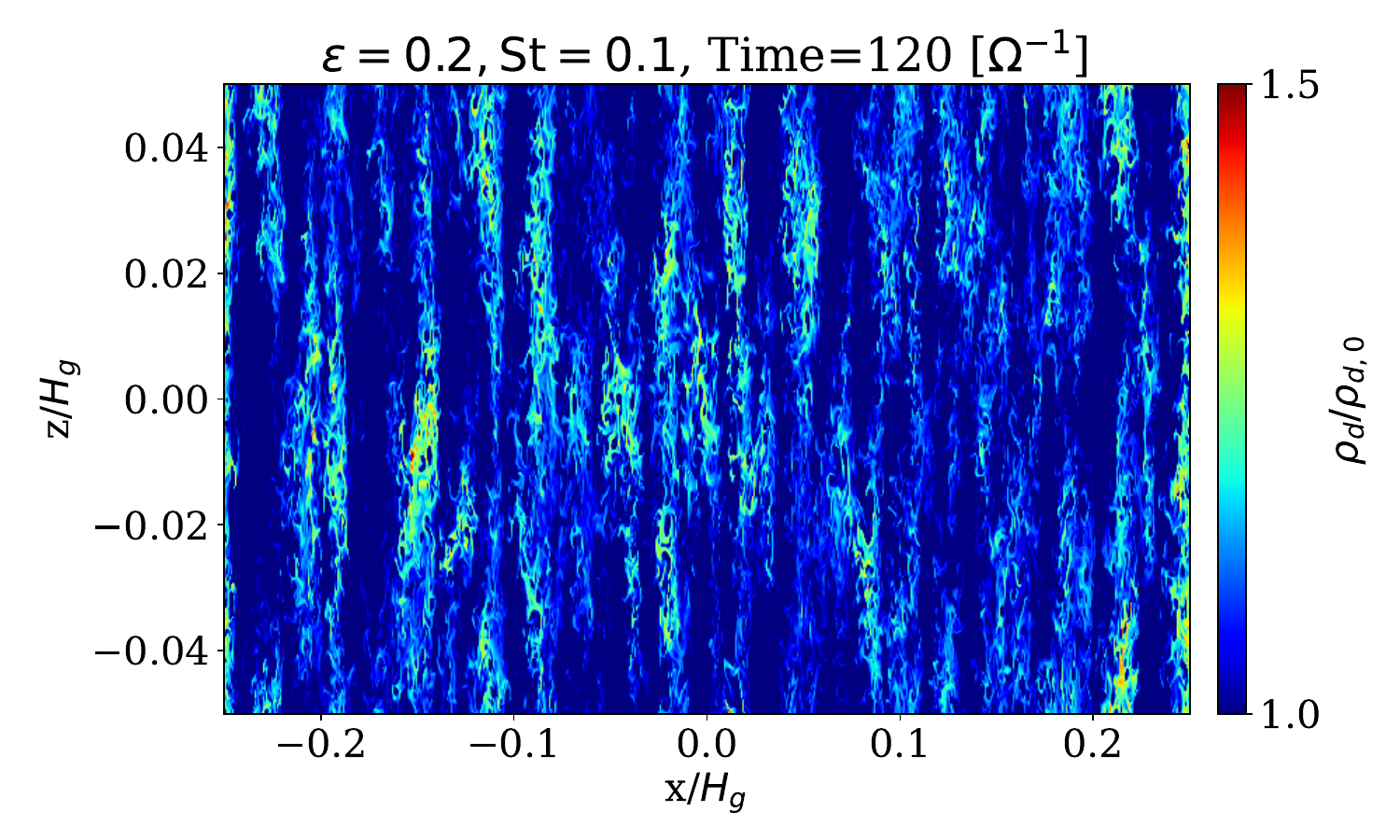}
\includegraphics[width=0.7\textwidth]{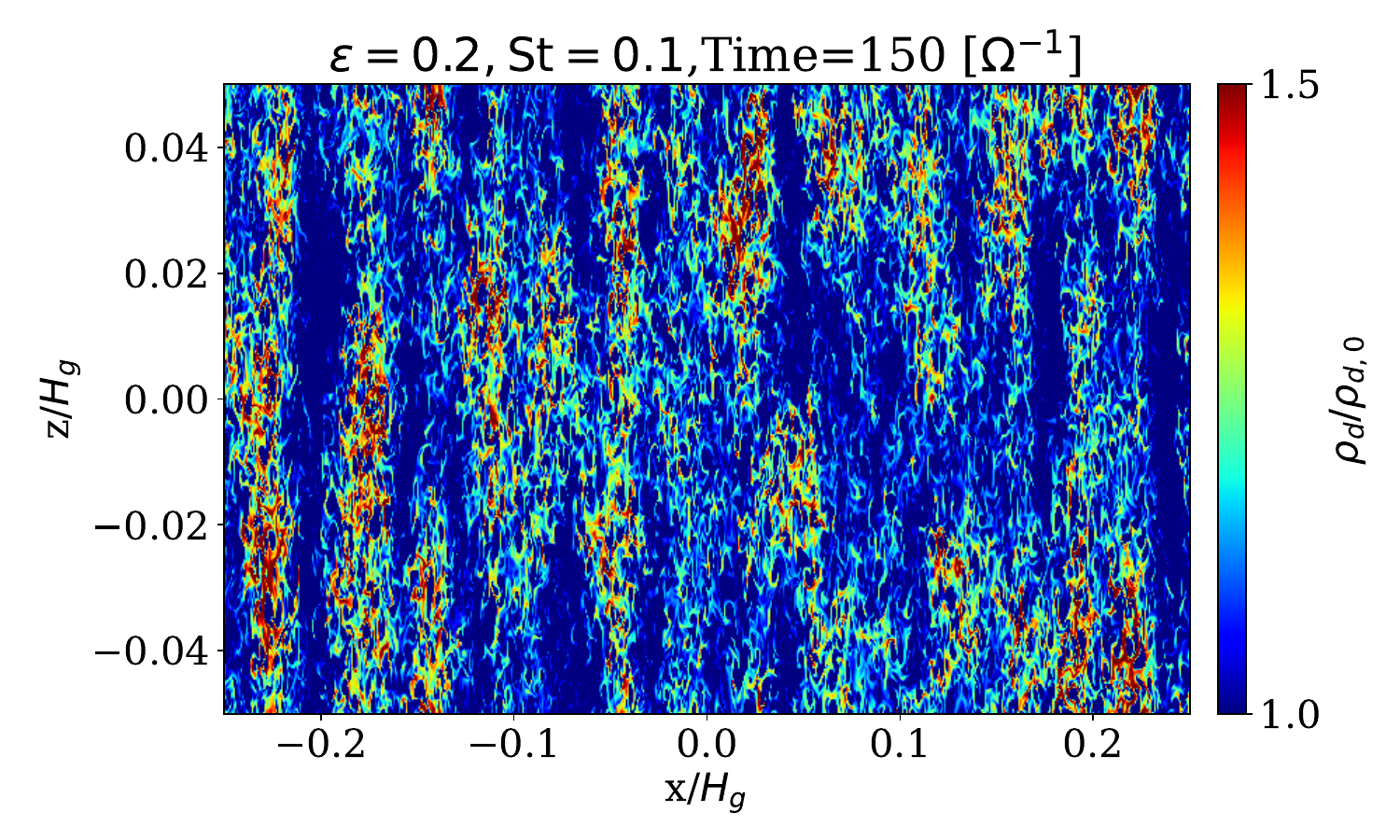}
\caption{For $\epsilon=0.2$ and $\st=0.1$ (run E0v2S0v1c, with coagulation), dust density snapshots at $t=120\;\Omega^{-1}$ (left) and $t=150\;\Omega^{-1}$ (right). }
\label{fig:st0v1t120}
\end{figure*}

\begin{figure}
\centering
\includegraphics[width=\columnwidth]{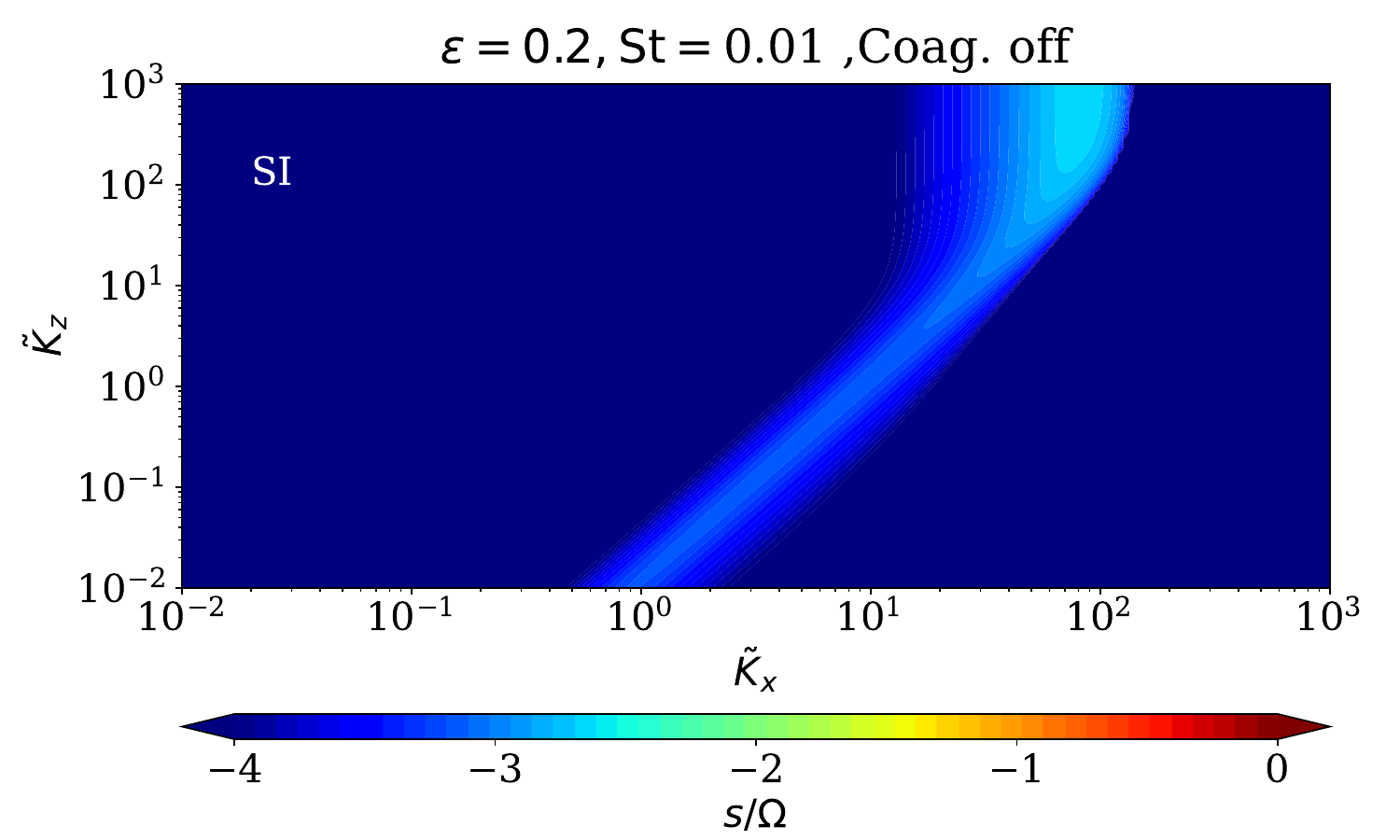}
\includegraphics[width=\columnwidth]{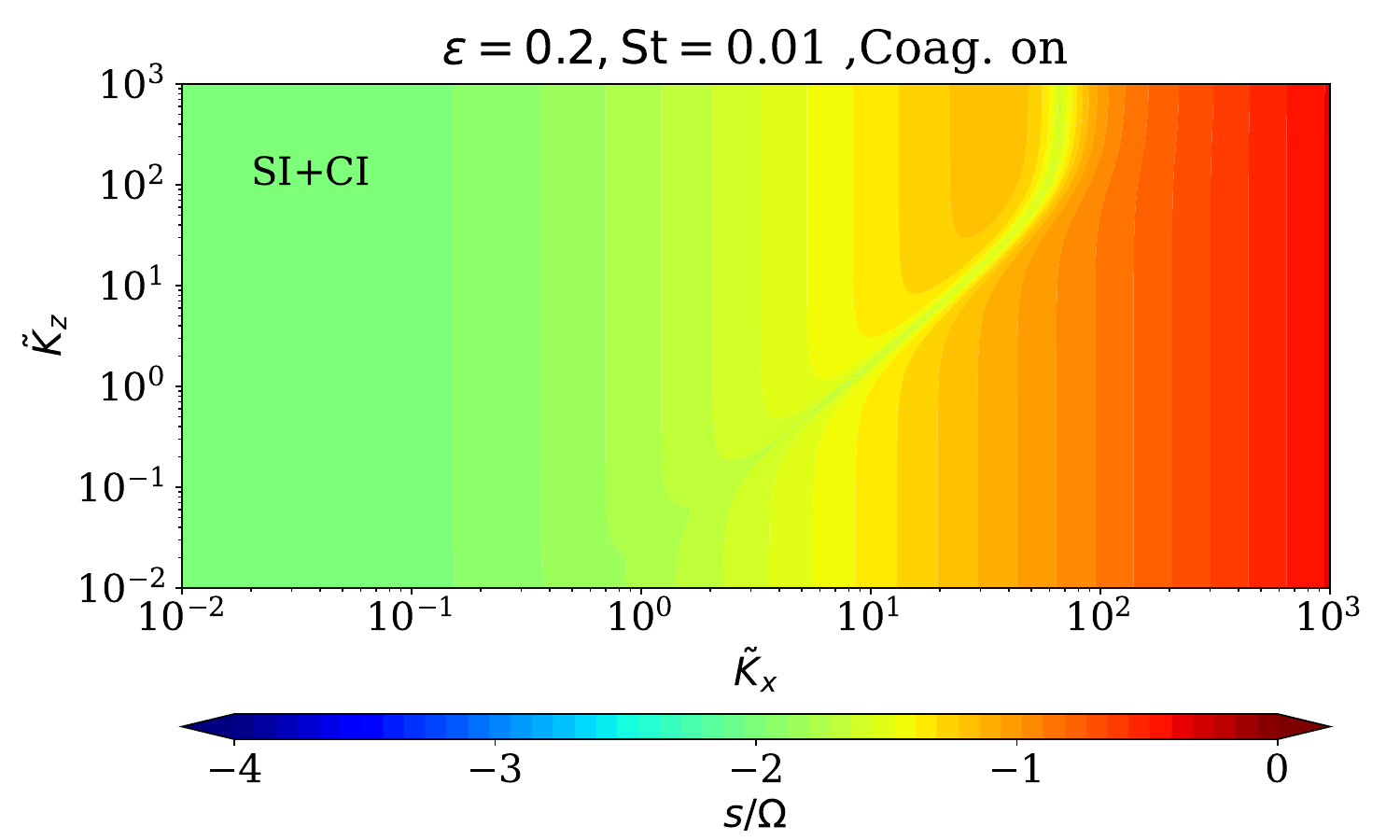}
\caption{Linear growth rates obtained from the linear analysis presented in Appendix \ref{sec:appA} for $\epsilon=0.2$ and $\st=0.01$ for the case without coagulation (top) and with coagulation(bottom). Here, $\tilde K_{x,z}=k_{x,z} \eta R_0$ are the dimensionless wavenumbers.}
\label{fig:SIassistCI}
\end{figure}

\begin{figure*}
\centering
\includegraphics[width=0.7\textwidth]{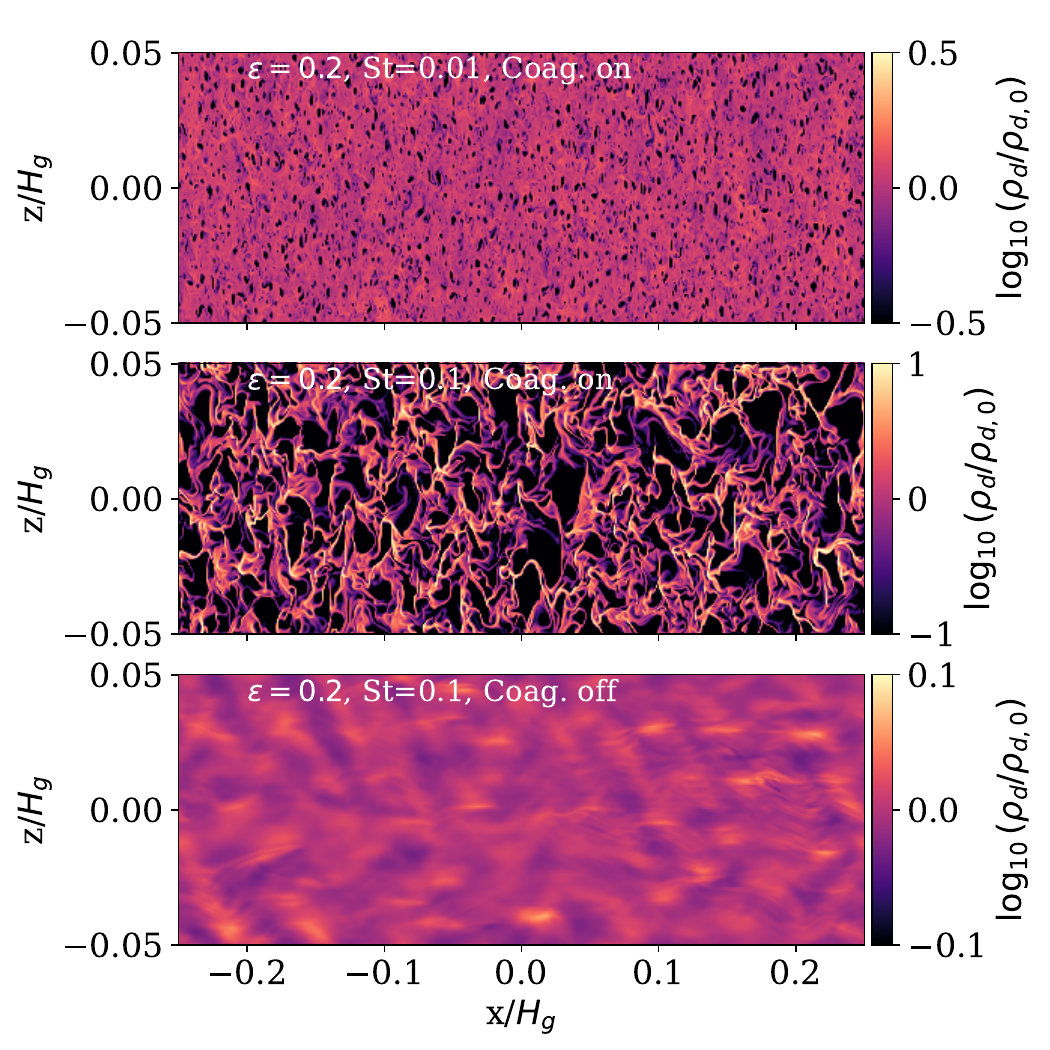}
\caption{From top to bottom: dust density map during the saturation phase for model E0v2S0v01c ($\epsilon=0.2$, $\st=0.01$, with coagulation), model E0v2S0v1c ($\epsilon=0.2$, $\st=0.1$, with coagulation), model  E0v2S0v1 ($\epsilon=0.2$, $\st=0.1$, without coagulation). }
\label{fig:isotropic}
\end{figure*}

\begin{figure*}
\centering
\includegraphics[width=\textwidth]{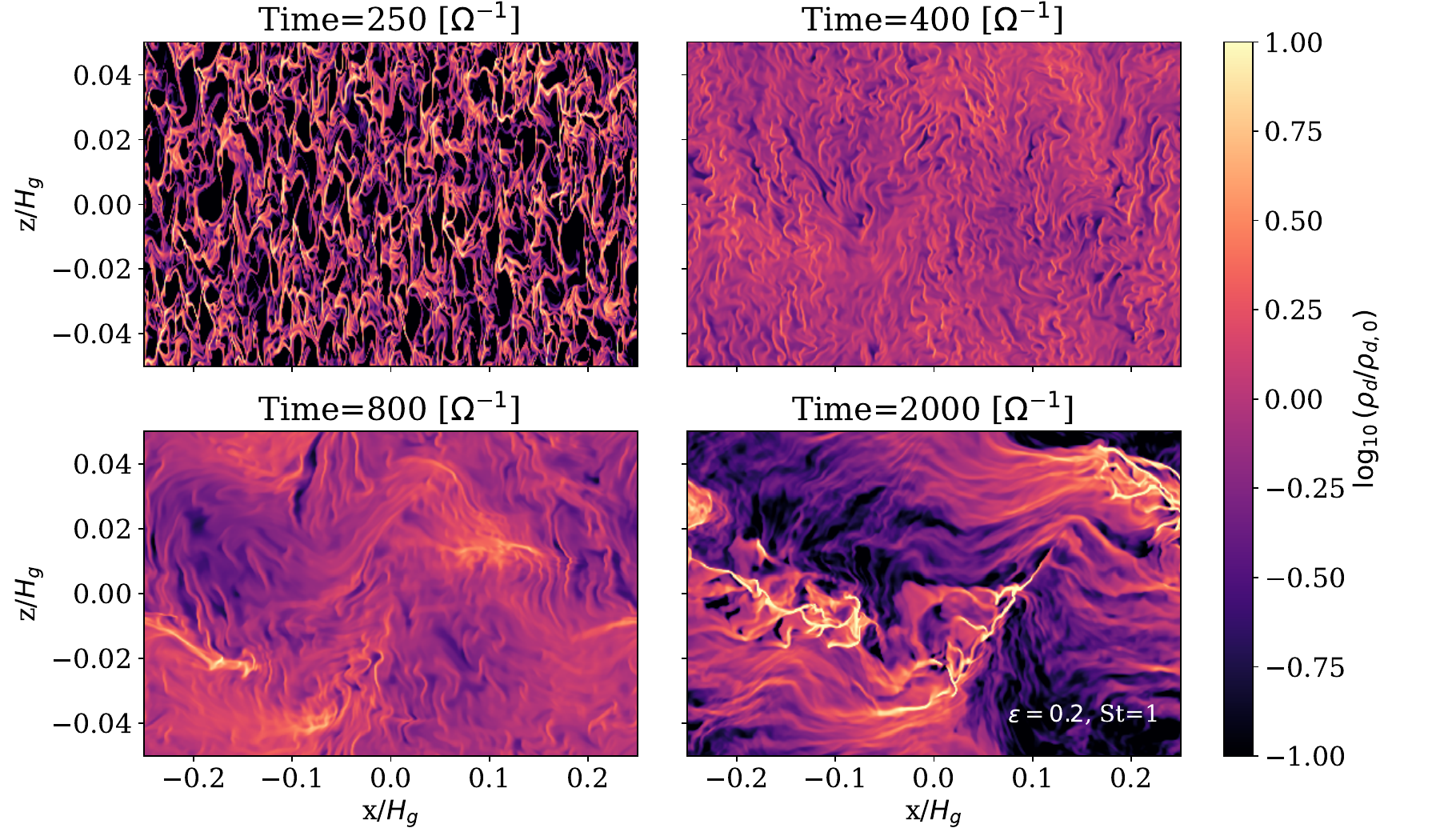}
\caption{For $\epsilon=0.2$ and $\st=1$ (model E0v2S1c, with coagulation),  dust density snapshots taken at different times. }
\label{fig:e0v2s1}
\end{figure*}

\begin{figure}
\centering
\includegraphics[width=\columnwidth]{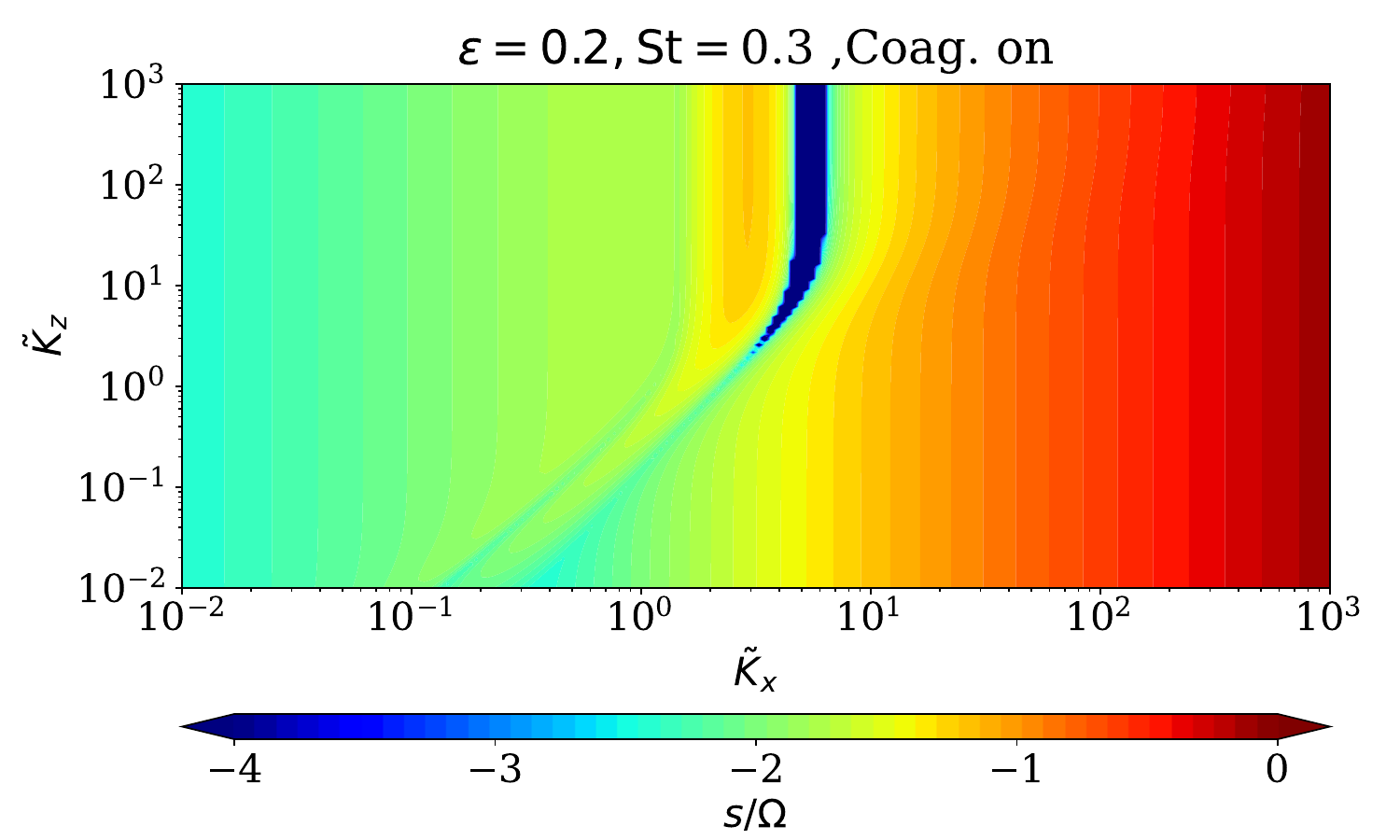}
\includegraphics[width=\columnwidth]{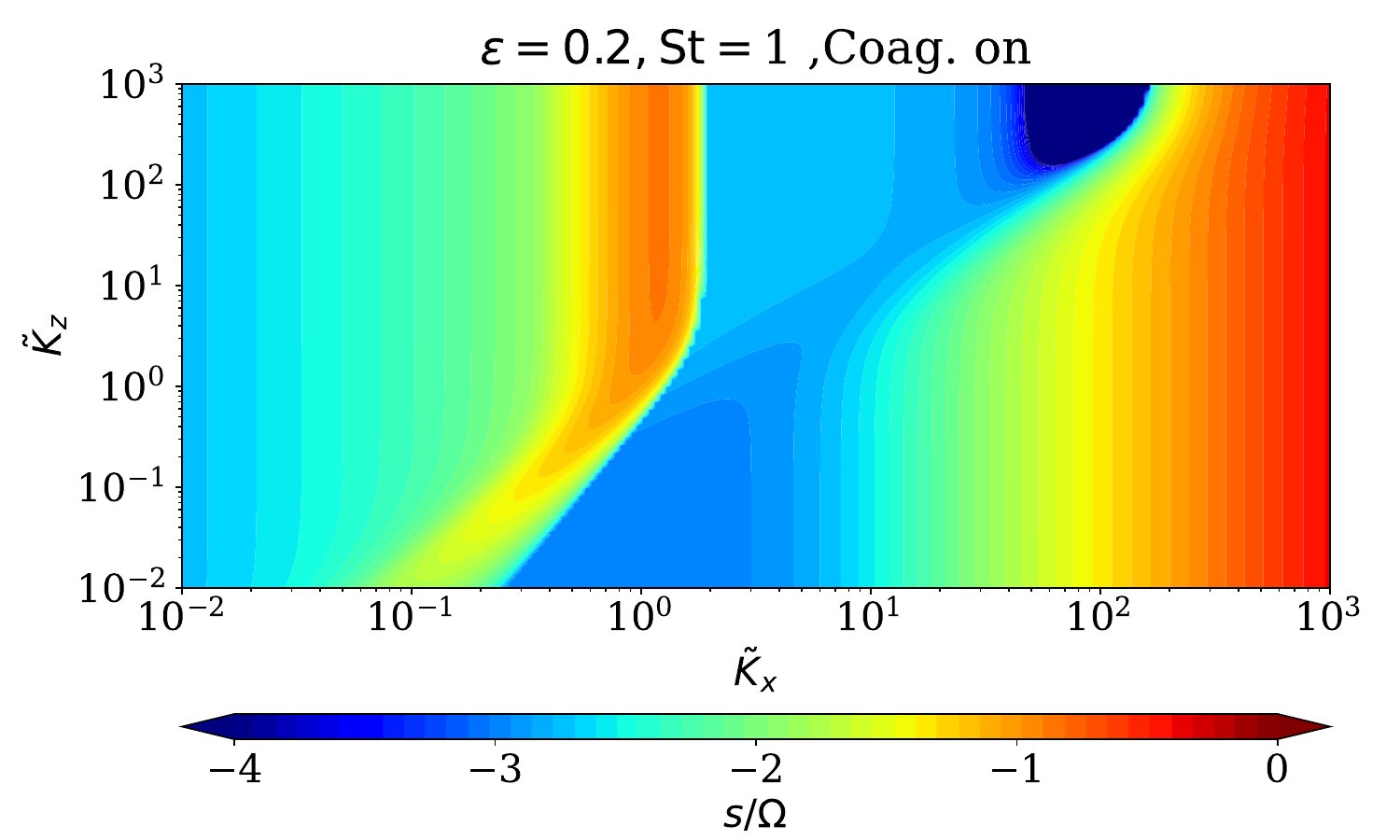}
\caption{For $\epsilon=0.2$ and with the effects of coagulation included, linear growth rates obtained from the linear analysis presented in Appendix \ref{sec:appA}  for $\st=0.3$ (top) and  $\st=1$  (bottom). Here, $\tilde K_{x,z}=k_{x,z} \eta R_0$ are the dimensionless wavenumbers.}
\label{fig:sicicompetition}
\end{figure}

In the case with $\epsilon=0.2$, we find  that the difference in amplitude between the $\alpha_{g,i}$ coefficients is smaller. For simulations including coagulation, their time evolution  is represented   in Fig. \ref{fig:alphagi_eps0v2}. The underlying turbulence is more isotropic in that case,  with $\alpha_{g,y}\sim \alpha_{g,z}$ for both $\st=0.01$  (Model E0v2S0v01c) and $\st=0.1$  (Model E0v2S0v1c).  This is also illustrated in Fig. \ref{fig:spectrum} where we plot the kinetic energy spectrum at  saturation for $\st=0.1$. The power spectrum is broadly consistent with a negative five third power law (dashed line), which suggests Kolmogorov-like turbulence.   Isotropic turbulence is expected for  classical SI-driven turbulence, although we find  an early  evolution driven by the CI. In Fig.  \ref{fig:dustmax}, the top-middle and middle-middle panels show the evolution for $\st=0.01$ and $\st=0.1$ respectively. For models with coagulation,  the initial rise in $\epsilon_{max}$ (for $t\lesssim$ 100 $\Omega^{-1}$ typically)   is caused by the CI, involving density growth of vertically extended filaments through dust capture and merging of these filaments into denser ones. Compared to  the previous case with $\epsilon=0.02$, however, it appears that the value for $\epsilon_{max}$ is  high enough to trigger the SI within the filaments, which leads to  the subsequent increase in $\epsilon_{max}$ that is observed. This would be consistent with the isotropic diffusion mentioned above. Fig. \ref{fig:st0v1t120} shows the dust density at the onset of  the secondary increase of $\epsilon_{max}$.  It is clear that non-linear overdensities can grow in very  localized regions within the filaments. This corresponds to unstable modes with vertical wavenumbers $k_z \ne 0$, which favors an SI-like instability .  To confirm that the SI is definitely at work inside the filaments, we conducted a test calculation by restarting the run for $\st=0.1$ at  $t=200\Omega^{-1}$, but with the dust back-reation onto the gas switched off. The finding of this model is that  overdensities are quickly damped, which confirms the important role of the dust feedback for the formation of clumps in the filaments.

 Interestingly, with respect to simulations without coagulation,  we can notice that the overdensities that are produced by the SI in presence of coagulation are stronger.  For $\st=0.1$ (middle-middle panel)  and $\st=0.01$ (top-middle panel), models with coagulation yield $\epsilon_{max}\approx 6$ and $\epsilon_{max}\approx 0.3-0.35$, respectively,  whereas the growth of the instability without coagulation is marginal.  We argue that for this range of parameters, the SI may be assisted by dust coagulation. To support this argument, we present in Fig. \ref{fig:SIassistCI} growth rates of the SI (upper panel) and those obtained from our linear analysis with coagulation included (bottom panel), for $\st=0.01$. For the run with $\st=0.1$  (Model E0v2S0v1c),  $\st=0.01$ corresponds roughly to the value for  which overdensities start to grow inside the filaments (see middle-middle panel in Fig. \ref{fig:dustmax}). On this plot, the red-coloured rectangular region for $\tilde K_x \gtrsim 100$, with $\tilde K_x=k_x\eta R_0$ the dimensionless wavenumber,   corresponds to unstable CI modes  as these  modes  have no vertical structure. The region enclosed between $10 \lesssim \tilde K_x \lesssim 100$ and $10  \lesssim \tilde K_z \lesssim 100$ can rather be identified as unstable SI modes. By  comparison with the upper panel, we see that these modes have larger growth rates when coagulation is taken into account,  which suggests that coagulation can amplify unstable SI modes.  From Fig. \ref{fig:st0v1t120}, we estimate the typical radial extent of a vertically extended filament to be $0.02 H_g$, which corresponds to $\tilde K_x \sim 15$. This is well inside this region of coagulation-assisted SI, confirming thereby that this mechanism can  operate here. 
 
For $\st=0.1$  (Model E0v2S0v1c), inspecting the two panels of Fig. \ref{fig:st0v1t120} reveals that the vertical filaments broaden with time, which is simply a consequence of turbulent diffusion generated by the coagulation-assisted SI. As the filaments spead out, they can subsequently  start to intersect each other, which leads to an isotropic turbulent  configuration as suggested by the middle panel of Fig. \ref{fig:isotropic}. Note that a similar evolution is found for $\st=0.01$,  for which we show in the upper panel of Fig. \ref{fig:isotropic} a snapshot of the dust density when the instability has saturated. 

Going back to the middle panel Fig. \ref{fig:isotropic},  it is interesting to note that the fully turbulent state consists of  large voids and narrow particle streams, which is similar to the patterns developed by the SI for strongly coupled particles \citep{2007ApJ...662..627J}.  For  standard SI, however, this regime is typically  found for $\epsilon \gtrsim 1$ whereas for $\epsilon=0.2$, dust density is left almost unchanged (see bottom panel of Fig. \ref{fig:isotropic}). This implies that coagulation makes this mode of nonlinear saturation of the SI occur at much lower dust-to-gas ratios.  This is not surprising as coagulation tends to increase radial drift speed by increasing the Stokes number, so that growing particles can fall more rapidly through the voids created by Poisson fluctuations \citep{2007ApJ...662..627J}. We note,  however,  that the dust density enhancement is limited to $\sim 30-40$ with coagulation included,  whereas for standard SI this mode of nonlinear saturation generally corresponds to  strong clumping  with  dust density enhancements $\gtrsim 100$.

\subsection{Marginally coupled particles}
For $\epsilon=0.02$, the evolution outcome of  the model with  final Stokes number $\st=1$ and including coagulation (run E0v02s1c, bottom left panel in Fig. \ref{fig:dustmax}) is  found to be very similar to that of $\st=0.1$ particles (run E0v02s0v1c), involving the formation of a vertically extended  filament inside which CI-induced turbulence operates. 

For $\epsilon=0.2$, an interesting issue that emerges from the previous discussion in Sect. \ref{sec:e0v2} is whether the  conditions  to reach  strong clumping are less stringent with coagulation.  To investigate this further, we now allow, in the case where coagulation is included, the Stokes number to increase up to $\st=1$. For standard SI, a model with $\epsilon=0.2$ is expected to lie in the strong clumping regime for $\st\sim 1$ and our main aim here is to examine if this threshold is diminished when coagulation is taken into account.   In  Fig. \ref{fig:dustmax},  models with $\epsilon=0.2$ and $\st=1$  correspond to the bottom middle panel where we compare the runs with and without coagulation. From this plot, it is immediately evident that:  i) the final outcome of the simulations does not depend on whether coagulation is included or not, and ii) including coagulation does not reduce the threshold in terms of Stokes number for the onset of strong clumping. Looking at the time evolution of  $\epsilon_{max}$ for the model with coagulation, we see that the initial behaviour is consistent with the results presented in the previous section, involving increase of $\epsilon_{max}$ as coagulation-assisted SI grows and develops  isotropic turbulence. At a later time $t\sim 200$ $\Omega^{-1}$ , however, $\epsilon_{max}$ is observed to decrease before increasing again as $\st \rightarrow 1$. From the time evolution of $\st$, we estimate the (mean) Stokes number associated with the observed drop of $\epsilon_{max}$ occuring at  $t\sim 200$ $\Omega^{-1}$ to be  $\st  \approx 0.3$. Interestingly, we find by running simulations of standard SI with different Stokes numbers that this value coincides approximately with the one above which significant growth of the SI occurs. To explore the origin of this decrease in $\epsilon_{max}$ in more details, we have  conducted a number of test calculations, by restarting the simulation with coagulation (run E0v2S1c) at $t\sim 200$ $\Omega^{-1}$ but with dust coagulation/feedback  switched on/off. The findings of this study can be summarised as follows: 
\begin{itemize}
\item  a restarted run with dust backreaction onto the gas switched off leads to a continuous increase in $\epsilon_{max}$ due to CI, which is the opposite behaviour to that found in model E0v2S1c. This demonstrates that dust feedback plays  an important role in the decrease in $\epsilon_{max}$ seen in run E0v2S1c.
\item  a restarted run with dust coagulation switched off results in the development of the SI into stripes aligned in the vertical direction but tilted in the radial direction, similarly to the patterns formed by the SI  in the case of large particles \citep{2007ApJ...662..627J}. When applied to run E0v2S1c, this shows that from $t\sim 200$ $\Omega^{-1}$ onward, the SI can indeed be triggered,  consistenly with the results of our simulations of standard SI with various Stokes numbers.  During the saturated state, we find $\epsilon_{max}\sim 0.5-1$ for this test calculation, which is also in agreement with our standard SI calculation with $\st=0.3$. 
\item a restarted run with coagulation {\it and} dust backreaction switched off results in a maintained quasi-stationary turbulent state, with $\epsilon_{max}$ remaining close to its initial value. 
\end{itemize}
Taken together, these results suggest that provided its growth rate is high enough,  the SI can counteract the effect of the CI such that the final turbulent state is similar to that found in pure SI calculations. This is  illustrated in Fig. \ref{fig:e0v2s1} which shows the evolution of the dust density for the model with coagulation included  (run E0v2S1c).  The large stripes tilted toward the vertical direction that are expected to develop during nonlinear saturation of the SI for $\st=1$ particles can easily be recognized.  For  the particular model with $\epsilon=0.2$ and $\st=1$,  examining  Table \ref{table} confirms that the values $\alphass$ and $\alpha_{g,i}$ do not  depend on whether or not coagulation is included. 

To clearly demonstrate the  finding that the SI can overcome the effects of the CI, we show in Fig. \ref{fig:sicicompetition} linear growth rates of the instability for $\epsilon=0.2$ and two different values of $\st$. Here, both the CI and SI operate, with  the unstable CI (resp. SI) modes corresponding to the unstable modes that have little (resp. strong)  dependence on $\tilde K_z$. We see that as coagulation proceeds and Stokes number transitions from $\st =0.3$ to $\st=1$, the maximum growth rate of the SI increases whereas unstable CI modes with radial wavenumbers  in the range $\tilde K_x\in[5-100]$ are damped. This possibly arises because both CI and SI are driven by the radial drift of dust, which acts as a  finite energy reservoir for the two instabilities such that any of these two instabilities can only grow at the expense of the other.

\section{Discussion}
\label{sec:discussion}
\subsection{Impact of coagulation on the Streaming Instability}
Although a model with $\epsilon=0.2$ and $\st=1$ leads to strong clumping, consistently with the results of \citet{2007ApJ...662..627J}, a necessary condition for particle clumping through the SI is generally approximated as $\epsilon \gtrsim 1$ in  the disc midplane, due to the strong increase in the SI linear growth rates at $\epsilon \approx 1$ for $\st \lesssim 1$. The right panels of Fig. \ref{fig:dustmax} show the effect of coagulation on the SI for $\epsilon=3$. The SI linear growth rate for $\st=0.01$ is $\sigma \sim \Omega^{-1}$ but this case only produced modest overdensities, whereas strong clumping occured for $\st \gtrsim 0.1$, as expected. Clearly, the effect of coagulation appears to be negligible for $\epsilon=3$, regardless the (final) value for $\st$. Combined with our results for $\epsilon=0.2$ and $\st=1$, this suggests that for the optimally dust sizes and dust-to-gas ratios that maximize the  SI growth rate, coagulation has an almost  null effect. In that case, this  means not only that i) coagulation has no impact on the SI linear growth rate but also that ii) the dust concentration at saturation is not affected by coagulation.   Physically, this arises simply when the coagulation timescale is longer that the SI growth timescale, or equivalently when the particule coagulation growth rate is  smaller than the SI growth rate. From Eq. \ref{eq:mp}, the inverse of the coagulation timescale is approximately given by:

\begin{equation}
t_{\rm coag}^{-1}=\frac{1}{2}\sqrt{\frac{\pi \alpha C}{8\st}}\epsilon. 
\label{eq:stperturb}
\end{equation}

For $\st=0.01$ this leads to $t_{\rm coag}^{-1}\sim 0.01 \Omega^{-1}$, which is indeed smaller that the aforementioned SI growth rate.  Starting from an initial Stokes number $\st=10^{-3}$, coagulation can of course promote SI by increasing the average dust size by 1-2 orders of magnitude, consistently with the results of \citet{2024ApJ...975L..34H}, but we find that once the SI is triggered, its dynamics dominate over coagulation effects.

\subsection{Effect of coagulation on turbulence and dust diffusion} 
\label{sec:turbulence}
Nevertheless, we have also shown that under certain conditions, coagulation and SI can act in concert. This is particularly true in dust-poor discs where the SI can grow in the  vertically extended filaments formed by the CI, leading to a regime of coagulation-assisted SI that ultimately results in  a level of turbulence much higher compared to standard SI  (see Sect. \ref{sec:e0v2}). For each model, the $\alpha_{SS}$ dimensionless parameter that quantities the radial flux of gas orbital momentum is listed in the first column of Table \ref{table}.  Most runs have $\alphass<0$, in agreement with previous estimations of the flux of angular momentum in SI-driven turbulent discs \citep{2007ApJ...662..627J,2022ApJ...937...55H}.  The fact that $\alphass$ is more negative when coagulation is taken into account suggests that the radial transport of angular momentum induced by the CI is also inward. This is not surprising since, similarly to the SI \citep{2007ApJ...662..627J},   CI is  also powered by a global pressure gradient. 

An interesting issue that needs to be investigated is how the collision velocity resulting from the turbulence compares with the fragmentation velocity $v_f$.  It can indeed not be excluded that the turbulence driven by the coagulation and streaming instabilities in turn hinders further coagulation by producing collision velocities that exceed the fragmentation threshold. To address this question, one can estimate the maximum Stokes number $\st_f$ achievable through coagulation. From Eq. \ref{eq:dvpp}, it is straightforward to show that $\st_f$ is given by: 
\begin{equation}
\st_f=\frac{v_f^2}{C \alphass c_s^2}
\end{equation}
The fragmentation velocity  has been studied both numerically and experimentally \citep{2008ARA&A..46...21B,2010A&A...513A..57Z,2013A&A...559A..62W,2015ApJ...798...34G,2019ApJ...871...10U}, and is estimated to be between 1 and 10 $m.s^{-1}$ for silicate particles and between 10 and 80 $m.s^{-1}$ for icy aggregates. Regardless the nature of dust aggregates, all but one simulations resulted in turbulent collision  velocities smaller than $v_f$ for typical disc parameters. Assuming that $R_0$ corresponds to 1 AU, Model E0v2S0v1c which enters the regime of coagulation-assisted SI is the only exception, as we find $\st_f \approx 0.07$ for $v_f=1m.s^{-1}$. By preventing the formation of large dust grains, this implies that  coagulation-driven turbulence may in that case inhibit the strong clumping phase of the SI. 

A related effect of coagulation is the vertical diffusion of particles driven by the CI and/or SI. At steady-state, namely once turbulent diffusion counterbalances gravitational settling, it is expected that the particle vertical profile is characterized by a dust scale height $H_d$ given by \citep{1995Icar..114..237D,2007Icar..192..588Y}:
\begin{equation}
\frac{H_d}{H_g}=\sqrt{\frac{\alpha_{g,z}}{\alpha_{g,z}+\st}}
\end{equation}

Using the values for $\alpha_{gz}$ reported in Table \ref{table}, we find that  it is possible to differentiate between CI and SI in dust-poor discs with $\epsilon\lesssim 0.2$ and for tightly coupled particles with $\st \lesssim 0.1$, as the SI is ineffective at producing a significant level of turbulence under these conditions.  By contrast, the CI on its own can generate a vertical bulk diffusion coefficient $\alpha_{gz}\sim 5 \times 10^{-7}$ for  $\epsilon=0.02$, which results in $H_d \sim 0.006 H_g$.  In the coagulation-assisted SI regime (run E0v2S0v1c), $\alpha_{gz}$ can even reach ${{\cal O }(10^{-4})}$ for $\st=0.1$, corresponding to $H_d \sim 0.01 H_g$.

\subsection{Caveats}
The results that we presented in this work may suffer from a number of assumptions and limitations,   the main one being that we adopted a moment equation to describe dust coagulation,  based on the vertically integrated Smoluchowski equation \citep{2016A&A...589A..15S,2021ApJ...923...34T}.  Rather than focusing on  the evolution of the full dust size distribution, this approach consists in following the evolution of the so-called peak mass which represents the weighted average mass. \citet{2016A&A...589A..15S} have shown that provided the effect of finite size dispersion is taken into account when evaluating the collision velocities, good agreement is obtained between single-size and full size calculations of drifting and growing dust grains in protoplanetary discs.  We caution, however, that the original model of \citet{2016A&A...589A..15S} does not take into account the effect of fragmentation and it is not clear whether or not the single-size approximation would still hold when fragmentation is considered, since this process would tend to broaden the initial dust size distribution.  Relaxing the single-size approximation and considering a full, possibly wide,   dust size distribution produced by fragmentation may have important consequences on our results. In the context of the SI, recent studies have indeed shown that taking into account the effect of a dust size distribution can significantly modify the growth rate of (monodisperse) SI \citep{2019ApJ...878L..30K,2020MNRAS.499.4223P,2021MNRAS.501..467Z,2021MNRAS.502.1469M}. Therefore, our model cannot explore the effect of coagulation on this "polydisperse" SI, and this should be examined in a future study. 

We have also considered inviscid discs, while various hydrodynamical instabilities can be triggered in the planet-forming regions of protoplanetary discs. This gives rise to turbulence, with a corresponding turbulent viscosity such that $\alphass\sim 10^{-4}$ typically. These levels of external turbulence may significantly impact the non-linear evolution of both the SI \citep{2020ApJ...891..132C}, and CI \citep{2021ApJ...923...34T}. Although models including dissipation effects should be explored in more details, we present in App. \ref{sec:appB} the results of a few viscous simulations that also take into account dust diffusion. These suggest that for the results presented in this paper may be affected by the external turbulence for $\alphass\gtrsim 10^{-6}$. 

Another limitation resides in the size of our vertical domain which, from the discussion in Sect. \ref{sec:turbulence}, is systematically higher than the dust layer thickness. The consequence is that the effects of stratification may be important and these should be examined in more details in a future study. Another reason for considering vertically stratified discs is that the CI is characterised by vertically extended filaments and it is not clear whether or not these would persist in stratified models. 

We note that  these two aspects, namely, considering a full dust size distribution and vertical stratification, have been recently studied by \citet{2024ApJ...975L..34H}. Although the potential important effect of the CI was not investigated in this work, they found that the dust size distribution can be significantly altered by coagulation within the dense dust clumps that are formed as the dust settles toward the midplane and the SI is triggered. 

\section{Summary}
In this paper, we have presented the results of unstratified, axisymmetric, shearing-box simulations of streaming instability (SI) including the effect of dust coagulation.  We solved a simplified version of the coagulation equation by a moment method, which consists in  following the size evolution of solids that dominate the total dust surface density. Our primary aim was to examine  to what extent dust growth  impacts the linear growth of the SI, and to get insight into its impact on the nonlinear saturation phase of the SI. We are particularly interested in the role played by the coagulation instability (CI, \citet{2021ApJ...923...34T}), which results from the dust-density dependence of coagulation efficiency and the grain size dependence of radial drift speed. Our motivation stems from the fact that the CI can be triggered for dust-to-gas ratios as small as $\epsilon \sim 10^{-3}$ and make the Stokes number of solids rapidly reach unity, leading thereby to optimal conditions for efficient growth of the SI.  We studied how the CI and the SI can interact with each other, and how the interplay between these two instabilities depends on the initial dust-to-gas ratio and on the maximum Stokes number $\st$ allowed before fragmentation  sets in. Our key findings are summarized as follows:
\begin{enumerate}
\item In dust-poor discs with $\epsilon\sim 10^{-2}$ and for tightly coupled particles with $\st \lesssim 10^{-1}$, we find that the evolution is mainly driven by the CI. Since the latter instability is by essence an  axisymmetric instability, early evolution involves the formation of vertically extended filaments that tend to merge themselves more rapidly for larger Stokes numbers. Within the filaments, nonlinear saturation of the CI gives rise to anisotropic turbulence with an enhanced dust diffusion coefficient in the vertical direction. 
\item In dust-rich discs with $\epsilon >  1$, the evolution is mainly driven by the SI simply because in that case, the growth timescale of the SI is smaller than the coagulation timescale. In this regime, coagulation is observed to have only little effect,  not only on the initial linear growth rate of the instability, but also on the final dust concentration values reached once the SI has saturated. 
\item For intermediate dust-to-gas ratios $\epsilon \sim 0.1$ and Stokes numbers $\st \lesssim 0.1$, we observe the onset of the SI within the filaments formed originally by the CI. Examination of linear growth rate maps indicates that for this range of parameters, the linear growth rate of the SI is significantly increased by coagulation effects, which we refer to as a regime of coagulation-assisted SI. In this regime, nonlinear evolution leads to isotropic turbulence with dust diffusion coefficients having similar value in each direction.  The generated turbulence causes the vertically extended filaments to spead out until they start to intersect each other, leading ultimately to a turbulent state characterised by large voids and narrow particle streams. The patterns that develop look similar to those formed by the SI in the strong clumping regime, albeit with a smaller dust concentration, of order ${\cal O}(30-40)$, compared to that typical of the strong clumping regime. 
\item As the Stokes number increases and approaches unity, however, SI can overcome the effect of the CI such that the final turbulent state is very similar to that found in pure SI calculations without the effect of coagulation included. 
\end{enumerate}
Taken together,  these findings suggest that the formation of dense clumps through the SI may be rendered easier by coagulation, simply because the process of dust growth enables the SI to reside in a regime where it can operate efficiently.  The CI by itself has only little impact on the onset of the SI, but it can nonetheless be a non negligible source of turbulence in protoplanetary discs, especially when it  can act in concert with the SI. 

\begin{acknowledgements} 
Computer time for this study was provided by the computing facilities MCIA (M\'esocentre de Calcul Intensif Aquitain) of the Universite de Bordeaux and by HPC resources of Cines under the allocation A0170406957 made by GENCI (Grand Equipement National de Calcul Intensif). MKL is supported by the National Science and Technology Council (grants 113-2112-M-001-036-, 114-2112-M-001-018-, 113-2124-M-002-003-, 114-2124-M-002-003-) and an Academia Sinica Career Development Award (AS-CDA-110-M06). This work was partly supported by the "Action Thématique de Physique Stellaire" (ATPS) of CNRS/INSU PN Astro co-funded by CEA and CNES , and by ORCHID program (project number: 49523PG, co-PI. Dutrey-Tang), funded by the French Ministry for Europe and Foreign Affairs, the French Ministry for Higher Education and Research and the National Science and Technology Council (NSTC). 
\end{acknowledgements}

\bibliographystyle{aa}
\bibliography{refs}
\begin{appendix}
\section{Linearized equations}
\label{sec:appA}
In order to linearize equations Eqs. \ref{eq:rhog}-\ref{eq:vdust}, we write $X$ as a quantity of the system as the sum between the equilibrium state $X_0$ (Eqs.  \ref{eq:uxgas}-\ref{eq:uydust}), and a perturbation $\delta X << X_0$ in Fourier modes with $k_x$ and $k_z$ the spatial modes and $\sigma$ the growth rate of the instability.
\begin{equation}
X(x,z,t) = X_0(x,z) + \delta X \times e^{i (k_x x + k_z z) + \sigma t}
\label{eq:perturbation}
\end{equation}

By injecting this into the mass continuity equations Eq.\ref{eq:rhog} and Eq.\ref{eq:rhod}, we obtain 
\begin{subequations}
    \begin{align}
    \sigma \delta \rho_g &= - \rho_{g, 0} (i k_x \delta u_x + i k_z \delta u_z) - i k_x u_{x, 0} \delta \rho_g \\
    \sigma \delta \rho_d &= 
    \begin{aligned}[t]
    & - \rho_{d, 0} (i k_x \delta v_x + i k_z \delta v_z) - i k_x v_{x, 0} \delta \rho_d \\
    & - \rho_{d, 0} D (k_x^2+k_z^2) (\delta \rho_d / \rho_{d,0} - \delta \rho_g / \rho_{g,0})
    \end{aligned}
    \label{eq:continuity SB gas+dust perturb}
    \end{align}
\end{subequations}

Similarly, with the equations of conservation of momentum for the gas Eq.\ref{eq:gasmom} 
\begin{subequations}
    \begin{align}
    \sigma \delta u_x &= -u_{x,0} i k_x \delta u_x + 2 \Omega \delta u_y - c_s^2 i k_x \frac{\delta \rho_g}{\rho_{g, 0}} + \delta F_{x}^{visc} \notag \\
    & \quad - C_{DF} m_p^{-1/3}
    \begin{aligned}[t]
    &\Bigg\{ [\delta \rho_d - \frac{1}{3 m_p} \rho_d \delta m_p] (u_{x,0} - v_{x,0}) \\
    &+ \rho_d (\delta u_x - \delta v_x) \Bigg\}
    \end{aligned}\\
    \sigma \delta u_y &= -u_{x,0} i k_x \delta u_y - \frac{1}{2} \Omega \delta u_x + \delta F_{y}^{visc} \notag \\
    & \quad - C_{DF} m_p^{-1/3}
    \begin{aligned}[t]
    & \Bigg\{ [\delta \rho_d - \frac{1}{3 m_p} \rho_d \delta m_p] (u_{y,0} - v_{y,0}) \\
    &+ \rho_d (\delta u_y - \delta v_y) \Bigg\}
    \end{aligned}\\
    \sigma \delta u_z &= -u_{x,0} i k_x \delta u_z - c_s^2 i k_z \frac{\delta \rho_g}{\rho_{g, 0}} + \delta F_{z}^{visc} \notag \\
    & \quad - C_{DF} m_p^{-1/3} \rho_d (\delta u_z - \delta v_z)
    \label{eq:momenta SB gas perturb x y z}
    \end{align}
\end{subequations}
with $C_{DF} = \Omega \sqrt{\frac{8}{\pi}} \frac{H_g}{\rho_i} (\frac{4 \pi \rho_i}{3})^{1/3}$ a coefficient related to drag force and $\delta F_{(x,y,z)}^{visc}$ the linearized viscous forces from \citet{2020ApJ...891..132C}:
\begin{subequations}
    \begin{align}
    \delta F_{x}^{visc} &= -\nu(k_z^2 + \frac{4}{3}k_x^2)\delta v_x - \frac{1}{3} \nu k_x k_z \delta v_z\\
    \delta F_{y}^{visc} &= - \nu (k_z^2+k_x^2) \delta v_y\\
    \delta F_{z}^{visc} &= -\nu(\frac{4}{3}k_z^2 + k_x^2)\delta v_z - \frac{1}{3} \nu k_x k_z \delta v_x
    \end{align}
    \label{eq:linearized viscous forces}
\end{subequations}

and for the equations of conservation of momentum for the dust Eq.\ref{eq:vdust},

\begin{subequations}
    \begin{align}
        \sigma \delta v_x &= -v_{x,0} i k_x \delta v_x + 2 \Omega \delta v_y \notag \\
        & \quad - C_{DF} m_p^{-1/3}
        \begin{aligned}[t]
        & \Bigg\{[\delta \rho_g - \frac{1}{3 m_p} \rho_g \delta m_p] (v_{x,0} - u_{x,0}) \\
        & + \rho_g (\delta v_x - \delta u_x) \Bigg\}
        \end{aligned}\\
        \sigma \delta v_y &= -v_{x,0} i k_x \delta v_y - \frac{1}{2} \Omega \delta v_x \notag \\
        & \quad - C_{DF} m_p^{-1/3}
        \begin{aligned}[t]
        &\Bigg\{ [\delta \rho_g - \frac{1}{3 m_p} \rho_g \delta m_p] (v_{y,0} - u_{y,0}) \\
        &+ \rho_g (\delta v_y - \delta u_y) \Bigg\}
        \end{aligned}\\
        \sigma \delta v_z &= -v_{x,0} i k_x \delta v_z - C_{DF} m_p^{-1/3} \rho_g (\delta v_z - \delta u_z)
        \label{eq:momenta SB dust perturb x y z}
    \end{align}
\end{subequations}

And finally, the perturbed equation for the coagulation equation Eq.\ref{eq:mp}

\begin{equation}
    \sigma \delta m_p = - v_{x,0} ik_x \delta m_p + D_{coag} m_p^{5/6} \rho_g^{-1/2} \rho_d 
    \left[ \frac{5}{6} \frac{\delta m_p}{m_p} - \frac{1}{2} \frac{\delta \rho_g}{\rho_g} + \frac{\delta \rho_d}{\rho_d} \right]
    \label{eq:coagulation perturb}
\end{equation}
with $D_{coag} = \left( \frac{\pi}{2} \right)^{5/12} \times 3^{5/6} \rho_i^{-1/3} \sqrt{2.3 \alpha} H_g^{-1/2} c_s$.\\

This system of 9 equations is rewritten in its matrix form as $s \mathbf{V} = A \mathbf{V}$, using the dimensionless quantities $\hat{v}=\frac{v}{\eta R \Omega}$, $K = k \eta R$ and $s = \frac{\sigma}{\Omega}$, with the perturbation vector 
$\mathbf{V}~=
\begin{bmatrix}
\frac{\delta \rho_g}{\rho_g},&
\frac{\delta \rho_d}{\rho_d},&
\delta \hat{u}_x,&
\delta \hat{u}_y,&
\delta \hat{u}_z,&
\delta \hat{v}_x,&
\delta \hat{v}_y,&
\delta \hat{v}_z,&
\frac{\delta m_p}{m_p}
\end{bmatrix}^\mathsf{T}$
and the matrix of the linearized system A.

\begin{figure}
\centering
\includegraphics[width=\columnwidth]{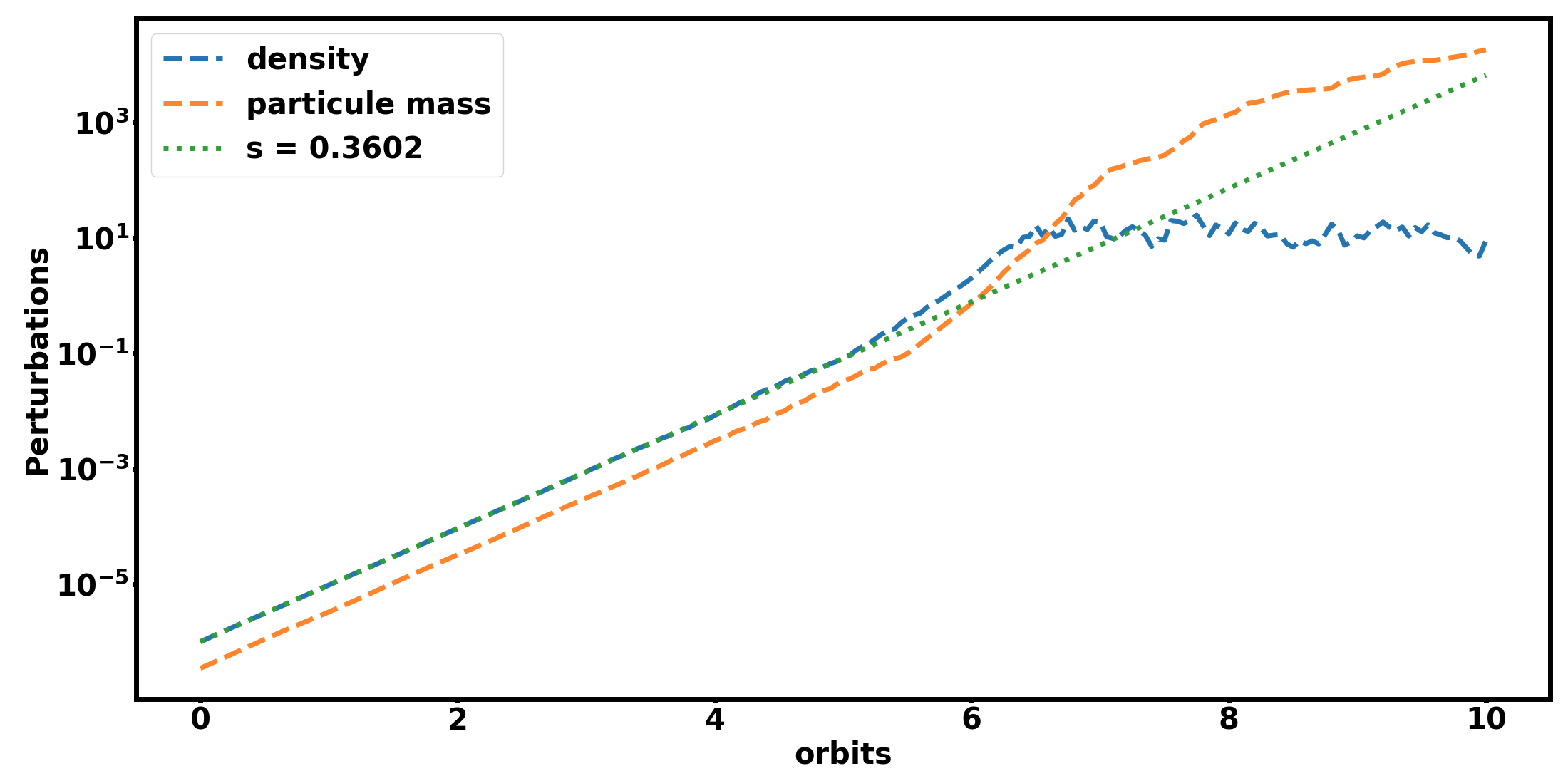}
\caption{In blue and orange, the perturbation $\frac{max(\lvert \rho_d - \rho_{d,0} \rvert )}{\rho_{d,0}}$ and $\frac{max(\lvert m_p - m_{p,0} \rvert )}{m_{p,0}}$ obtained through hydrodynamic simulations with coagulation. In green, the theoretical growth rate obtained analytically.}
\label{fig:appA_growthrate}
\end{figure}

For a given spatial mode ($K_x$, $K_z$), the largest eigenvalue of $A$ corresponds to the growth rate $\sigma$ of the most unstable mode, while the associated eigenvector can be used to study the linear case of this mode numerically. This allows verification of the coagulation implementation as well as the analytical development growth rates derived from the matrix's eigenvalues by comparing the two, done Figure~\ref{fig:appA_growthrate} in the linA case $St=0.1$, $\epsilon = 3$ and $(K_x, K_z) = (30,30)$ with coagulation. The analytical solution gives $s = 0.3602$, identical to the numerical growth rate derived from the particle density and particle mass until the perturbation is no longer negligible ($\delta X \sim 0.1$).

\section{Effect of gas viscosity and dust diffusion}
\label{sec:appB}
\begin{figure}
\centering
\includegraphics[width=0.91\columnwidth]{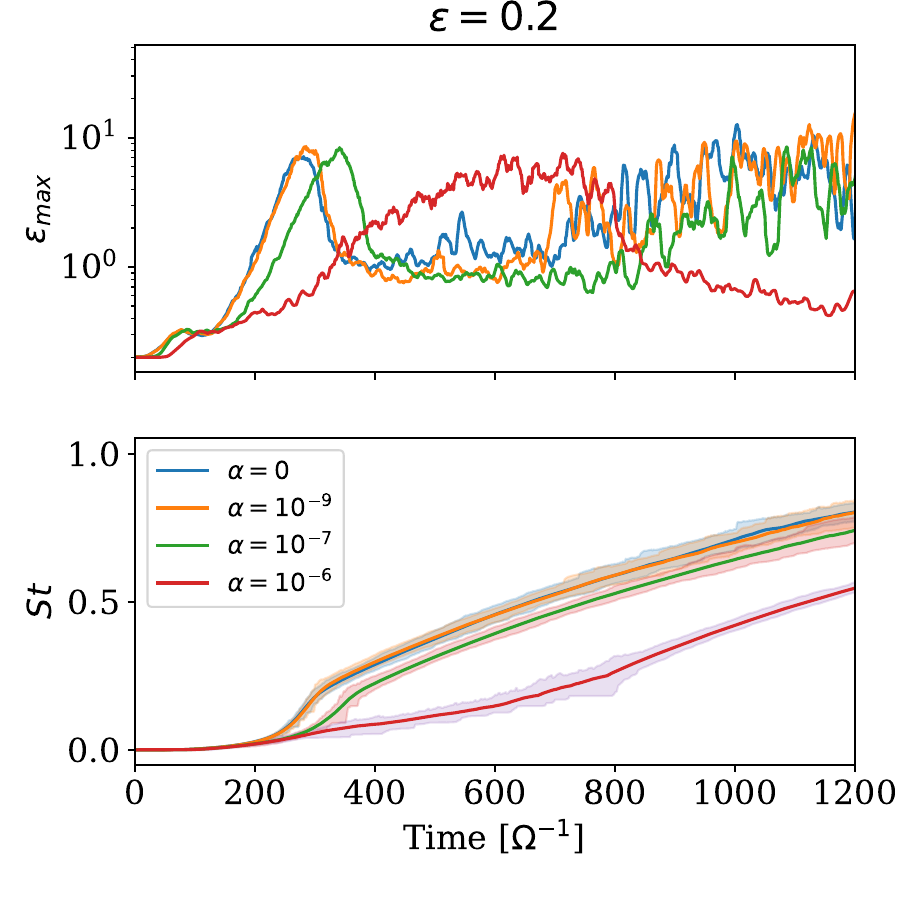}
\caption{For $\epsilon=0.2$, $\st=1$ and with coagulation included, time evolution of  the maximum dust-to-gas ratio $\epsilon_{max}$ and Stokes number $\st$  in viscous simulations where the $\alpha$ viscous stress parameter is taken to be equal to the dimensionless dust diffusion coefficient. }
\label{fig:viscous_evolve}
\end{figure}

\begin{figure}
\centering
\includegraphics[width=\columnwidth]{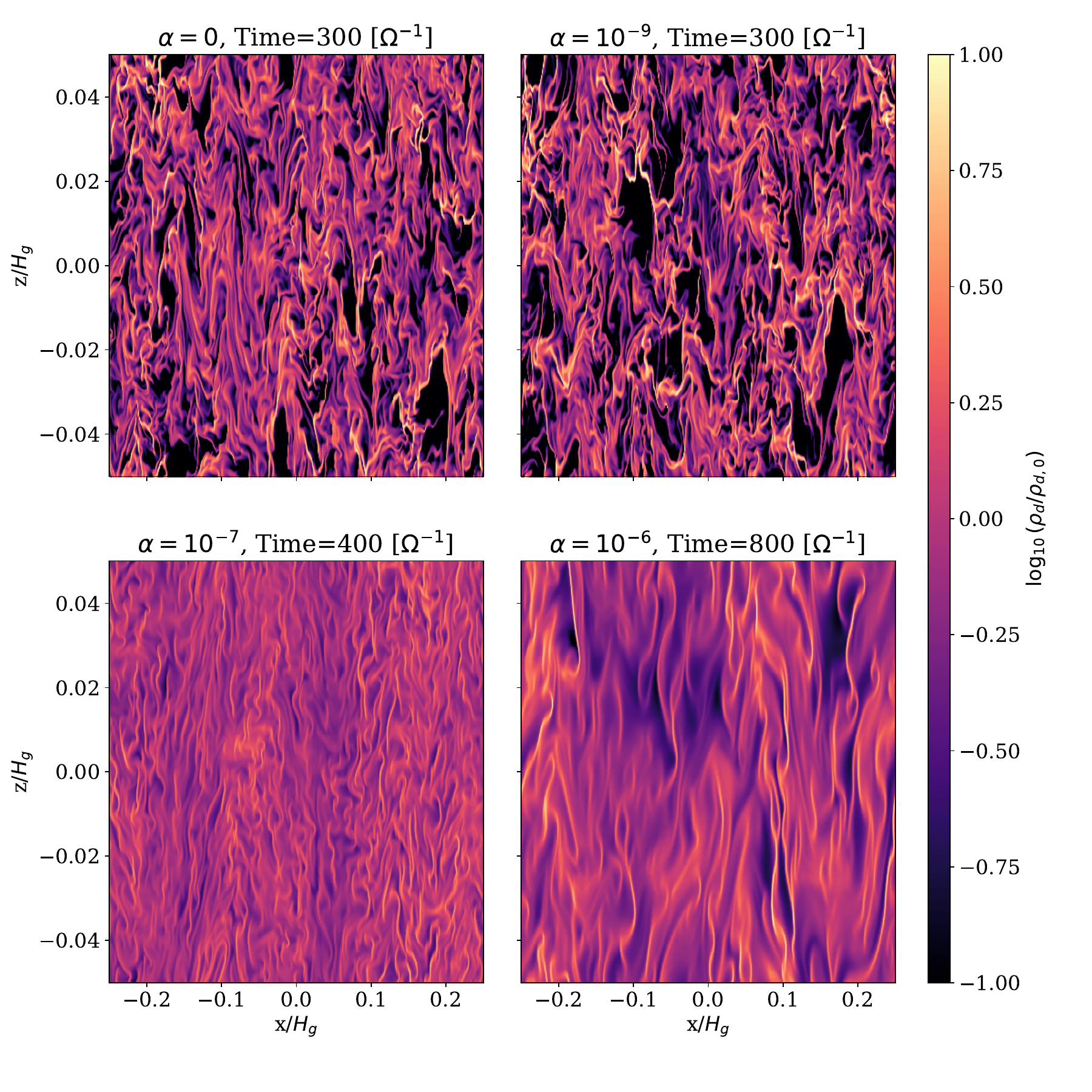}
\caption{For $\epsilon=0.2$, $\st=1$, dust density snapshots for the reference inviscid run (top left) and for viscous simulations with $\alpha=10^{-9}$ (top right),  $\alpha=10^{-7}$ (bottom left), and $\alpha=10^{-6}$ (bottom right). }
\label{fig:viscous}
\end{figure}

In the inviscid limit that we considered in this work, the numerical dissipation may impact our results, such that these could be dependent on the employed  resolution. This is because the growth rate of CI increases with radial wavenumber, such that the maximum wavenumber that is resolved by the grid is expected to be unstable. Here, we present the results of additional simulations with non-zero gas viscosity,  parameterized as $\nu=\alpha c_SH_g$, and with dust diffusion coefficient set to the same value $D=\nu$. As in inviscid simulations presented above, the $\alpha_{\rm coag}$ parameter entering in Eq. \ref{eq:dvpp} is set to $\alpha_{\rm coag}=10^{-4}$.  The aim is to examine i) how  the results are changed when dissipation processes are controlled and ii)  how these would be modified in presence of external turbulence, namely when the turbulence originates from other sources of instabilities than the CI and/or SI themselves. For different values of standard $\alpha$ parameter, the time evolution of $\epsilon_{max}$ and $st$ for these runs is presented in Fig. \ref{fig:viscous_evolve}. Not surprisingly, the linear growth rate of the instabilities is reduced as $\alpha$ is increased.  This is a consequence of the stabilization of the coagulation instability by dust diffusion \citep{2021ApJ...923...34T}. Unstable high wavenumbers modes are also suppressed, as  illustrated by Fig. \ref{fig:viscous} which shows the dust density at the end of the linear growth phase.  From Fig. \ref{fig:viscous_evolve}, we estimate the value for the numerical viscosity to correspond to $\alpha\sim 10^{-9}-10^{-7}$.  This is in line with the expectation that unstable modes with wavenumbers such that $kH \sim 1/\sqrt{\alpha}$ are damped by turbulence (Dubrulle et al. 1995).  Using $\alpha=10^{-7}$, this gives $k H \approx 3\times 10^{3}$, which is close to the value mentioned in Sect. \ref{sec:411}. Going back to Fig. \ref{fig:viscous_evolve}, we observe growth of the SI for each value of $\alpha$ we considered, except for  $\alpha=10^{-6}$. In that case, however,  $\st \lesssim 0.5$ at the end of the simulation due to slower coagulation, leading to a reduced SI growth rate. As $\st \rightarrow 1$, the SI should nevertheless enter a strong clumping regime, as it is expected in that case the SI growth rate to be only impacted by diffusion for $\alpha =10^{-6}$ (e.g. Figure 3. of \citet{2020ApJ...891..132C}, upper left panel). 

\end{appendix}

\end{document}